\documentclass{emulateapj}
\usepackage{apjfonts}
\usepackage{epsfig}
\usepackage{psfig}
\usepackage{graphicx}
\usepackage{natbib}
\usepackage{epstopdf}

\newcommand{\etal}{{\em et~al.}}

\def\aj{{AJ}}

\def\annrev{{ARA\&A}}
\def\apj{{ApJ}}
\def\apjs{{ApJS}}

\def\lax{{$\mathrel{\hbox{\rlap{\hbox{\lower4pt\hbox{$\sim$}}}\hbox{$<$}}}$}}
\def\gax{{$\mathrel{\hbox{\rlap{\hbox{\lower4pt\hbox{$\sim$}}}\hbox{$>$}}}$}}
\def\simlt{\lower.5ex\hbox{$\; \buildrel < \over \sim \;$}}
\def\simgt{\lower.5ex\hbox{$\; \buildrel > \over \sim \;$}}

\def\mnras{{MNRAS}}

\def\percm2{cm$^{-2}$}

\shorttitle{\textit{Chandra} Observations of 3C\,449}
\shortauthors{Lal et~al.}

\begin{document}

\title{Gas Sloshing and Radio Galaxy Dynamics in the Core of the 3C\,449 Group}

\stepcounter{footnote}

\author{Dharam V. Lal$^{1}$; Ralph P. Kraft$^{1}$;
Scott W. Randall$^{1}$; William R. Forman$^{1}$; Paul E.~J. Nulsen$^{1}$;
Elke Roediger$^{2}$; John A. ZuHone$^{3}$;
Martin J. Hardcastle$^{4}$;
Christine Jones$^{1}$; Judith H. Croston$^{5}$}
\affil{$^{1}$Harvard-Smithsonian Center for Astrophysics, 60 Garden Street, Cambridge, MA 02138, USA}
\affil{$^{2}$Jacobs University Bremen, PO Box 750 561, 28725 Bremen, Germany}
\affil{$^{3}$NASA-GSFC, 8800 Greenbelt Rd., Greenbelt, MD 20771}
\affil{$^{4}$University of Hertfordshire. School of Physics, Astronomy, and Mathematics, Hatfield AL 10 9AB, UK}
\affil{$^{5}$University of Southampton, School of Physics and Astronomy, Southampton, SO17 1SJ U.K.}

\begin{abstract}
We present results from a 140 ks \textit{Chandra}/ACIS-S observation
of the hot gas around the canonical FR I radio galaxy 3C 449.
An earlier, shorter 30~ks \textit{Chandra} observation of the group gas
showed an unusual entropy distribution and a surface brightness edge in the 
gas that could be a strong shock around the inner radio lobes.
In our deeper data we find no evidence for a temperature increase inside of the 
brightness edge, but a temperature decrease across part of the edge. This suggests that the edge is  
a ``sloshing'' cold front due to a merger
within the last $\lesssim$1.3--1.6 Gyrs.
Both the northern and southern inner jets are bent slightly to
the west in projection as they enter their respective lobes,
suggesting that the sloshing core is moving to the east.
The straight inner jet flares at approximately the position
where it crosses the contact edge,
suggesting that the jet is entraining and thermalizing some of
the hot gas as it crosses the edge.
We also detect filaments of X-ray emission around the southern inner radio jet and lobe
which we attribute to low entropy entrained gas.
The lobe flaring and gas entrainment were originally 
predicted in simulations of \citet{loken95} and are confirmed in our deep observation.
\end{abstract}



\section{Introduction}
\label{sec:intro}

Groups and clusters of galaxies in the local Universe continuously undergo
mergers with small groups and galaxies falling into the larger potential.
Even if the mass ratio of the infalling object to that of the cluster is
relatively small, the merging process can have a significant impact on the
ICM of the larger body.  In particular, a minor merger 
can offset the gas core from the center of the dark matter potential.
The offset gas core will oscillate around the center of the dark matter potential
creating a characteristic spiral pattern in the temperature and density structure
\citep{yago06}.  This characteristic spiral pattern, referred to
as ``sloshing'', is commonly seen in Chandra observations of groups \citep{machacek11} and 
clusters of galaxies \citep{maxim07}.
The gas temperature and density change discontinuously at the sloshing front,
but the pressure typically remains at or near equilibrium across the front.
Hydrodynamic simulations of the process demonstrate that
subsonic flows are created parallel to the sloshing fronts \citep{yago06}.

Massive elliptical galaxies lie at the centers of every cluster of galaxies, and most or all of these
galaxies contain supermassive black holes which have outbursts and drive jets.
The understanding of astrophysical jets requires detailed knowledge of the hydrodynamic parameters
of the material both of the collimated flow and of the external medium.
The \textit{Chandra} X-ray Observatory (\textit{Chandra}) has opened a new window into the
study of extragalactic radio jets with its ability to spatially resolve X-ray
emission from the active nuclei, non-thermal emission from ultra-relativistic particles
accelerated in the jets and lobes created at internal shocks in the jets,
and thermal emission from the galaxy/group/cluster environments that confine the
radio-emitting plasma.
To date, non-thermal X-ray emission from more than 100 extragalactic radio jets
and hot spots has been reported \citep[][and references therein]{massaro11}, and the gaseous environments
of several hundred additional radio galaxies spanning a wide range of both jet power
and environmental richness have been studied in detail,
{\it e.g.}, M\,87 \citep{forman2005,forman07}, Hydra\,A \citep{nulsen2005},
MS\,0735.6$+$7421 \citep{mcnamara05}, Centaurus~A \citep{kraft07a,mjh03},
3C\,33 \citep{kraft07b}, 3C\,288 \citep{Laletal2010}, among many others.
The visible appearance of a radio galaxy depends in detail on the complex interplay between
the jet power, the internal Mach number of the jet, the external gas pressure and pressure profile, and the
velocity field in the external medium.

The two types of radio galaxies where the morphology is most influenced by
the external medium are the narrow angle tailed (NAT) and the wide angle tailed (WAT) radio galaxies.
NATs, such as 3C\,129 \citep{LalRao2004}, are
radio galaxies in which the twin jets are sharply bent in the same direction
and are believed to be the result of the infall of the jet host galaxy into
a larger potential \citep{jaffe73}.  WATs, on the other hand, such as 3C\,465
\citep{mjh05}, typically reside in group
and cluster cores and show more gentle bending of the radio jets.
How this jet bending relates to gas motions in the external medium remains uncertain.

The radio source 3C\,449 is a canonical example of an
Fanaroff-Riley class I radio galaxy \citep{fr74} and exhibits a complex radio
morphology on spatial scales from parsecs to hundreds of kilo-parsecs.
The source is characterized by an unresolved core,
two symmetrically opposed jets, and very extended lobes \citep{perley79}.
The jets are relativistic near their bases and decelerate significantly
within 10$^{\prime\prime}$ (= 3.3~kpc) from the core.
Each jet is remarkably straight for
$\sim$50$^{\prime\prime}$ ($\simeq$ 16.7~kpc)
from the nucleus before deviating sharply towards the west,
dropping in surface brightness and terminating in large diffuse lobes \citep{feretti99}.
Since both jets bend in the same direction, rotation effects cannot
cause this morphology \citep{birkinshaw81}.
On larger scales, the outer lobes \citep{feretti99} are diffuse
and are relaxed in appearance, although the southern lobe is spherical while
the northern lobe appears to be elongated.

The hot gaseous atmosphere around
3C\,449 has been observed with \textit{ROSAT},
\textit{XMM}-\textit{Newton}, and \textit{Chandra}.
\textit{ROSAT} observations \citep{mjh98} showed that
the group gas is not spherically symmetric, and that
southern lobe is embedded in a rim of hot gas, with a deficit at the
position of the radio lobe suggesting that a cavity has been created by
the inflation of the lobe.  No X-ray emission was detected in the vicinity of
the northern outer lobe, suggesting a significant external pressure difference between
the northern and southern outer lobes.
More recently, \citet{croston03}, using \textit{XMM}-\textit{Newton}
observations measured the temperature and density profile of the
ICM to a distance of $\sim$100 kpc from the nucleus and argued that the outer lobes of 3C\,449
are significantly older than expected from spectral age estimates if they
are evolving buoyantly.
They also showed that the gas temperature of the
group gas was significantly higher than that predicted from
the temperature-luminosity relationship.
Interestingly, \citet{croston03} also found that the temperature decreased
radially outward from the core, perhaps suggesting that the core
had been heated by the nuclear outburst

\citet{sun09}, using a 30~ks archival \textit{Chandra} observation,
reported a significant peak in the entropy profile of the 3C\,449 cluster gas,
consistent with the temperature gradient found by
\citet{croston03} and supporting their scenario of AGN heating.
The level of heating and the unusual entropy profile are
indicative of supersonic inflation of the radio lobes at fairly high
Mach number, $M$ $\sim$2 \citep{sun09}.
If the inflation of the inner lobes is driving a strong shock
into the group gas then the current outburst
is completely detached from the larger scale bubbles (i.e. that the outer lobes
are no longer receiving energy from the nucleus and are evolving buoyantly).
Re-examination of this archival \textit{Chandra}
observation indicates the presence of a surface brightness edge in the
gas (at low statistical significance) at roughly the distance
of the inner radio lobes.
Shocks have been detected around the radio lobes of a number of radio galaxies,
but most of these are relatively weak with $M$ $<$1.5.
Shock heating of the intracluster medium (ICM)
is thought to play a key role in the energy budget of the gas cores of clusters and
groups, and in the suppression of cooling flows.  Shocks with
$M>$1.5 are particularly important because they are
relatively rare and the equivalent heat input due to the entropy increase in the gas is $\sim$10\%.
Here, we present results from a new
deep \textit{Chandra} observation of the hot gas atmosphere around 3C\,449
to better understand the dynamics of the radio lobe/ICM interaction.

This paper is organized as follows:
Section~2 contains a summary of the observational details.
An overview of the data and observed structures are presented
in Section~3.
Results of the data analysis are presented in Section~4 and
we discuss their interpretation and implications in Section~5.
Section~6 contains a brief summary and conclusions.
A \textit{Wilkinson Microwave Anisotropy Probe} cosmology
with $H_0$ = 73 km s$^{-1}$ Mpc$^{-1}$, $\Omega_M$ = 0.27,
and $\Omega_\Lambda$ = 0.73 is adopted \citep{spergel07}.
At a redshift of $z$ = 0.017085 of the host galaxy of 3C\,449,
the look-back time is 2.26$\times$10$^{8}$ yr,
the angular size scale is 20.6 kpc arcmin$^{-1}$, and
the luminosity distance, $D_L$ is 71.1 Mpc.
All coordinates are J2000.
The elemental abundances that we quote are relative to the Solar value
tabulated by \citet{anders89}.  Absorption by gas in our
galaxy \citep[$N_{\rm H}$ = 1.19 $\times$ 10$^{21}$ cm$^{-2}$][]{dickey90}
is included in all our spectral fits.
All spectral analysis errors are 90\% confidence, while all other errors are
68\% confidence.
North is up and east is to the left in all images.
Throughout the paper, we define the spectral index $\alpha$ in the sense
that $S_{\nu}\propto \nu ^{-\alpha}$.

\section{Data Processing} \label{observation}

\begin{figure*}
\begin{center}
\begin{tabular}{l}
\includegraphics[width=16.0cm]{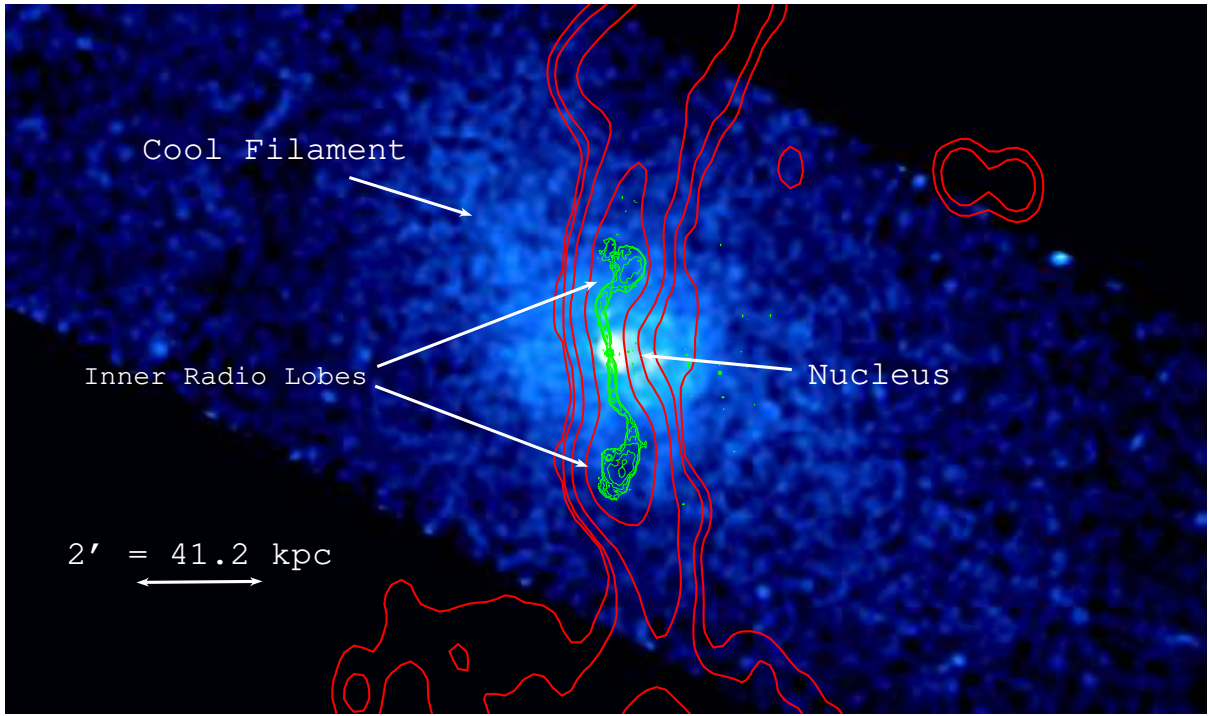}
\end{tabular}
\end{center}
\caption{
Combined
\textit{Chandra}/ACIS-S image of 3C\,449 in the
0.5--2.0 keV band.
The image is binned to 2$^{\prime\prime}$ pixels and smoothed
with a Gaussian ($\sigma$=4$^{\prime\prime}$.
All point sources, other than the active nucleus of 3C\,449, have been removed.
We see diffuse thermal emission in the soft band extending over
the entire field of view (12'$\sim$240 kpc along the east-west axis).
The green and red contours are taken from the 8.4 (7.1$''$ Gaussian beam)
and 0.6 GHz (47.9$''$ Gaussian beam) radio maps, respectively.
}
\label{counts_image}
\end{figure*}

3C\,449 was observed with \textit{Chandra} using ACIS-S.
We summarize the dates, exposure times
and \textit{Chandra} ObsID's in Table~\ref{chandra_obs}.
The data were reprocessed starting with the level~1
event files and using {\sc CIAO} version 4.3 with the up to date gain and
calibration files applied (CalDB version 4.4.1).
All observations were performed in Very Faint data mode.
Observation specific bad pixel files were produced and applied
and events flagged with {\it ASCA} grades 1, 5 and 7 were excluded.
Standard procedures were followed to exclude high background; i.e.,
for each observation, point sources and the region containing the brightest diffuse
group emission were excluded, and a light curve was extracted in the
0.3--10~keV range.
We excluded time intervals when the count rate deviated by more
than 20\% from the mean.
The cleaned exposure times
for each ObsID are listed in Table~\ref{chandra_obs} with the total cleaned
exposure time being 141.5~ks.

\begin{figure*}[ht]
\begin{center}
\begin{tabular}{l}
\includegraphics[width=12cm]{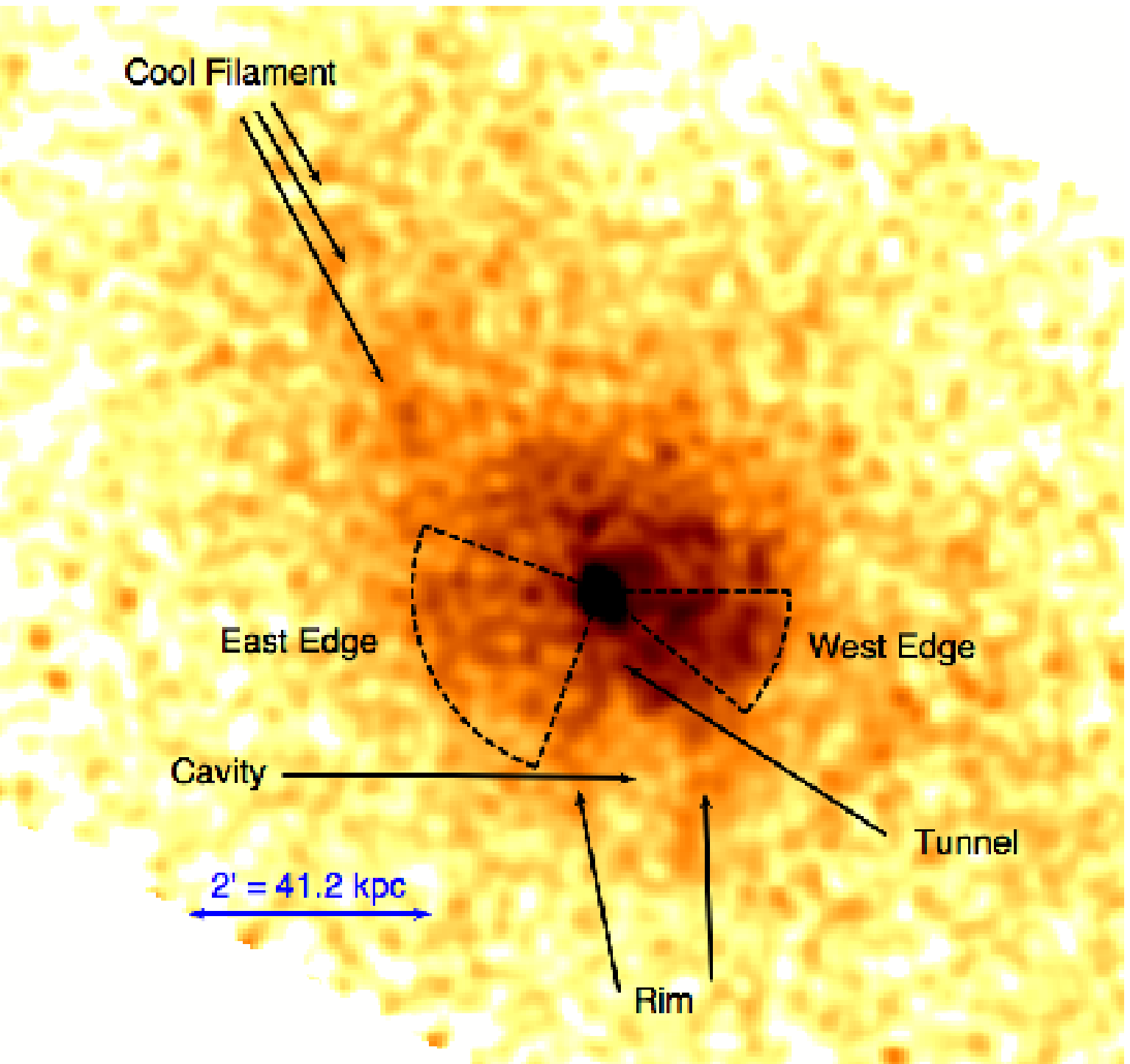}
\end{tabular}
\end{center}
\caption{{\small Background-subtracted, exposure-corrected, Gaussian-smoothed
($\sigma$=4$''$) \textit{Chandra}
image in the energy range 0.5-2.0~keV.
Features of interest which are discussed in the text are labeled including the cool
filament, the eastern and western edges, the rims and cavity associated with the southern
inner radio lobe, and the tunnel associated with the southern inner jet.  The two wedges
denote the regions for which the surface brightness profiles in Figures~\ref{se_sb_profile} 
and~\ref{w_sb_profile} were extracted.}}
\label{images}
\end{figure*}

\begin{table}
  \caption{\emph{Chandra} observations analysed in this paper.}
  \begin{tabular}{rllcc}
     \hline
    ObsID & Detector & ObsDate & Exposure & Cleaned Exposure \\
     &  &  & (ks) & (ks) \\ \hline
    4057 & ACIS-S & 2003-09-18 & 29.6 & 29.18 \\
    11737 & ACIS-S & 2010-09-14 & 53.0 & 52.35 \\
    13123 & ACIS-S & 2010-06-20 & 60.7 & 59.92 \\ \hline
  \end{tabular}
  \label{chandra_obs}
\end{table}

\subsection{Background Modeling} \label{back_mod}

Since the 3C\,449 group extends well beyond the Chandra
field-of-view, it was necessary to use
an external data set for background subtraction.
For background subtraction during spectral and imaging analyses,
we use the blank sky
observations\footnote{See http://cxc.harvard.edu/contrib/maxim/acisbg}
appropriate for the chip and epoch of the observation of interest. The
background files were processed in the same manner as the observations,
using the same background filtering, bad pixel and gain files, and were
reprojected to match the observations.
The source and background images were normalized in the 10--12~keV range.  We checked
for excess soft Galactic X-ray emission in the observations by extracting
spectra from source-free regions far off-axis (away from the group
emission) and compared to spectra extracted from the
dark sky background files.  We found no significant differences.

\subsection{Radio Data}

We use an archival WSRT image of 3C\,449 at 0.61~GHz \citep{perley79}
taken from the online 3CRR Atlas\footnote{ATLAS catalog:
Radio images and other data for the nearest 85 DRAGNs
(radio galaxies and related objects) in the so-called ``3CRR''
sample of \citet{laing83}.  Available at
{\tt http://www.jb.man.ac.uk/atlas/index.html} .}
which provides well-calibrated, well-sampled images.
We also make use of data from the VLA archive at 1.4, 4.9 and 8.4~GHz.
Details regarding the radio observations
are summarized in Table~1 of \citet{feretti99}.  The VLA data were reduced
in {\sc AIPS} (Astronomical image processing system) following the standard
procedures.  Absolute flux density calibration was tied to the observations
of a suitable flux calibrator close to the VLA archive data.

\section{Overview of X-ray data} \label{data_a}

Figure~\ref{counts_image} shows a Gaussian-smoothed, 
background-subtracted, exposure corrected image
of 3C 449 from the \textit{Chandra}/ACIS-S observations in the
0.5--2.0~keV band with point sources removed.
Radio contours from the WSRT 0.61 GHz map (red) and the VLA 8.4 GHz map
(green) are overlaid to show the relationship between the hot gas and
the radio plasma.  To enhance the visibility of structures in the diffuse emission we created
an additional zoom-in of the central region with features of interest
labeled; this is shown in Figure~\ref{images}.
These two images highlight the following features:
\begin{enumerate}
%
%
\item 
There is an elongated diffuse structure extending to
$\sim$330$^{\prime\prime}$ ($\simeq$110~kpc) to the northeast of the core, labeled 
``cool filament''.
\item There are two arc-shaped tangential edges in the surface brightness $\sim$100$^{\prime\prime}$
($\simeq$33~kpc) to the southeast and to the west of the core.
\item A close-up view of the jet in the southern direction shows
the presence of a ``tunnel-like'' feature, marked  with a dashed line.
The tunnel connects the core and the southern radio lobe and
is filled with the radio jet.
\item There is an X-ray cavity, i.e. a decrement in X-ray emission,  at the location of
the southern inner radio lobe.  This cavity is surrounded by a rim of enhanced brightness.
No cavity is detected at the position of the
northern inner radio lobe. 
\end{enumerate}
In this paper we present a detailed spatial and spectral analysis of these features
to constrain the gas dynamics of the system and better understand the interaction of the
hot gas with the radio jets.

\section{Data Analysis}

\subsection{Large-scale properties}

\begin{figure}
\begin{center}
\begin{tabular}{l}
\includegraphics[width=8.5cm]{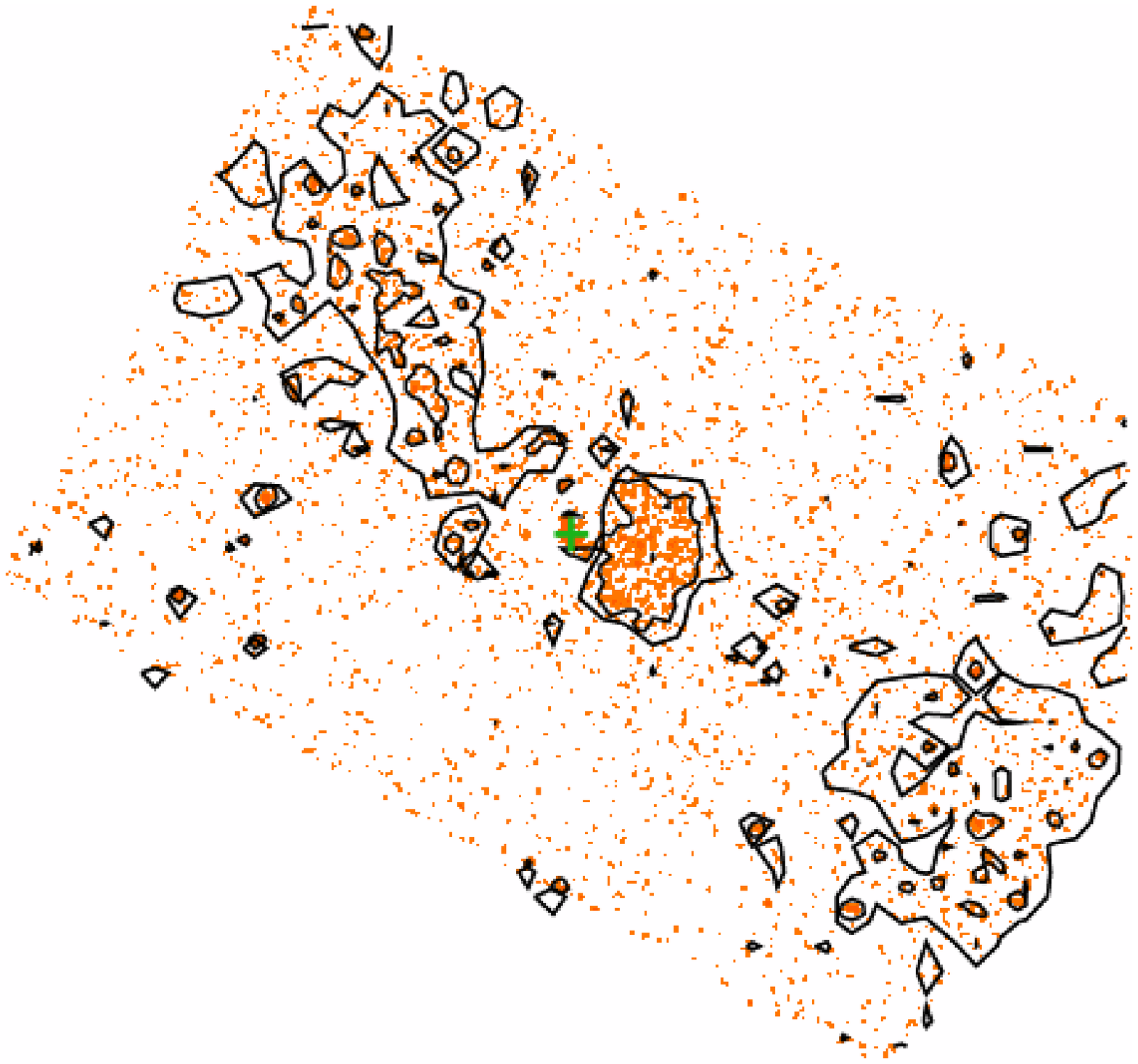}
\end{tabular}
\end{center}
\caption{{\small Residual image produced by subtracting an
azimuthally symmetric $\beta$ model,
This image is smoothed using a Gaussian kernel with radius of
3$^{\prime\prime}$. The contours are placed at 1.22 $\times$ 10$^{-6}$
and 1.77 $\times$ 10$^{-7}$ ~counts~s$^{-1}$~arcsec$^{-2}$.
The cool filament and an asymmetric core emission are clearly seen.
The green cross-point at the center shows the location of the radio/X-ray core.
}}
\label{residual}
\end{figure}

\begin{figure}[h]
\begin{center}
\begin{tabular}{l}
\includegraphics[width=6.0cm,angle=-90]{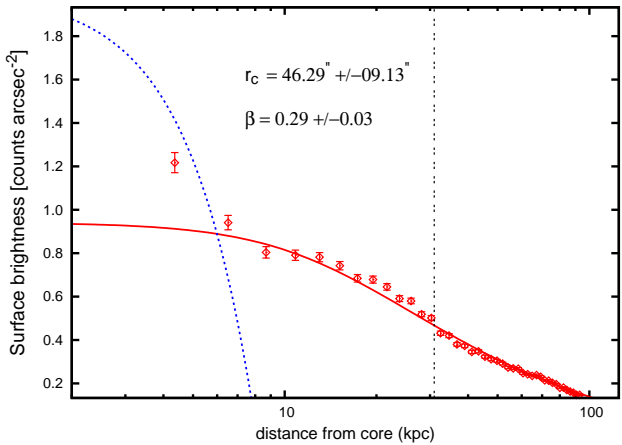}
\end{tabular}
\end{center}
\caption{{\small Azimuthally averaged 0.5--2.0~keV
radial surface brightness profile of 3C\,449
group gas centered on the active nucleus 
created from the combined \textit{Chandra} data.
The model is the best-fitting point-source plus $\beta$-model,
with $\beta$ = 0.29 and r$_{\rm c}$ = 46.29$^{\prime\prime}$ (= 15.41~kpc).
The vertical line shows the approximate location of the surface brightness
edge.}}
\label{beta}
\end{figure}

To characterize the large-scale spatial distribution of the X-ray-emitting gas,
we derived the radial surface-brightness profile of the ICM.
The profile was fitted with an azimuthally symmetric $\beta$-model
shown in Figure~\ref{beta}.
We found a best-fitting model with $\beta$ = 0.29 $\pm$0.03,
core radius r$_c$ = 46.29$^{\prime\prime}$ $\pm$9.13$^{\prime\prime}$
(= 15.41 $\pm$3.03 kpc)
with $\chi^2_\nu$ (DOF) = 1.32 (35) (confidence ranges are 1$\sigma$).
The core radius is roughly consistent within uncertainties with
the best-fitting model of \citet{croston03} derived from the XMM-Newton
observation, but the power law index is somewhat flatter.  The XMM-Newton
data covered a much larger solid angle on the sky, and there are clearly
non-azimuthally symmetric structures in the gas, so the exact fit parameters
will depend on exactly what distances from the nuclei are fit.
To emphasize the asymmetric structures revealed in
Figures~\ref{counts_image} and \ref{images} we created a residual image
by subtracting the average $\beta$-profile from the data.
The residual image in Figure~\ref{residual}
shows that the northeast filament and
an X-ray enhancement to the west of the nucleus are detected with high ($>$99\%) 
significance.
The western edge of the enhancement to the west is the position of one of the
surface brightness edges.
A substructure 5.6$^{\prime}$ (= 112 kpc) to the southwest is also visible.

\subsection{Large-scale Thermodynamic Structure}

We fitted a single temperature
{\sc VAPEC}\footnote{Astrophysical Plasmas Emission Code
- http://cxc.harvard.edu/atomdb} model
after excising point sources and the central AGN to the gas emission
in the energy range 0.5--5.0 keV,
with the absorption column density set at the
Galactic value, N$_H$ = 1.19 $\times$ 10$^{21}$ cm$^{-2}$
\citep{dickey90}.
The abundances for O, Fe and Si were free to vary,
with the rest of the elemental abundances (to which the spectral
fits are insensitive) fixed at 0.5~$\times$ Solar and
the temperature was free to vary.
The goal of this large scale fit is to constrain the elemental abundances
from a single spectral fit over a large region with a large number of counts.
The single temperature model
provided a good fit with $kT$ = 1.66$^{+0.04}_{-0.05}$~keV for $\chi^2$ (DOF) 0.98 (171),
and the abundances for O, Fe and Si were found to be
0.12$^{+0.21}_{-0.12}$, 0.32$^{+0.07}_{-0.06}$ and 0.56$^{+0.17}_{-0.16}$
$\times$~solar respectively.
The sub-Solar value of the Fe and Si abundance, and the extremely low
value of the O abundance, are consistent with X-ray observations of a sample
of early-type galaxies \citep{humphrey06}.  
We use these values of the elemental abundance for all subsequent spectral fits and
in the creation of the temperature maps.

\begin{figure}[ht]
\begin{center}
\begin{tabular}{l}
\includegraphics[width=5.5cm,angle=-90]{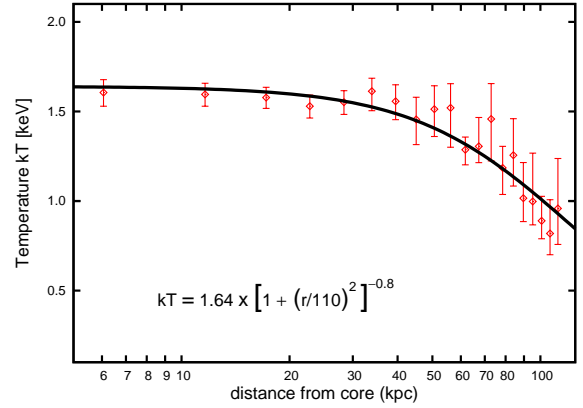}
\end{tabular}
\end{center}
\caption{{\small Plot of temperature variation in the environment of 3C\,449
as a function of distance from the core for azimuthally averaged profiles;
the black-line is the best-fitting analytic $kT(r)$ model.}}
\label{temp_rad_az}
\end{figure}

We fit the temperature of the gas in 20 radial bins from $\sim$2~kpc to $\sim$120~kpc.
An analytical approximation to the best-fitting values of $kT(r)$ over the
range of radii is given below:
$$
{kT} = 1.64 \times \left[1 + \left(\frac{r}{110} \right)^2 \right]^{-0.8} {\rm keV}.
$$
This approximation is shown in Figure~\ref{temp_rad_az} by the black solid line.
From this temperature profile,
we also find that the temperature decreases with increasing radius,
which is consistent with the results of \citet{croston03}.
The average temperature in the wedge
containing the cool filament to the northeast
(between the P.A. interval 10$^\circ$ and 70$^\circ$)
shows a similar change but the overall temperature is
lower than the azimuthally averaged temperature profile.

\begin{figure}[b]
\begin{center}
\begin{tabular}{l}
\includegraphics[width=5.5cm,angle=-90]{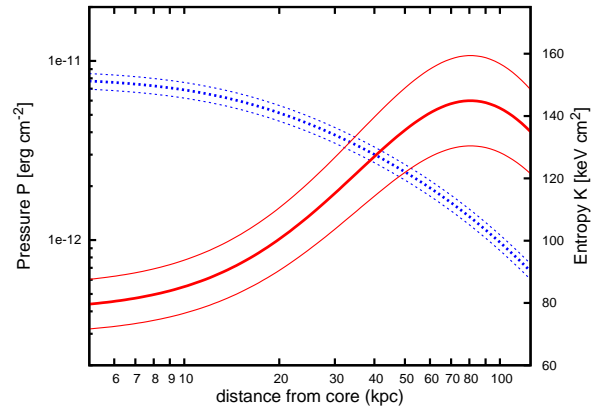}
\end{tabular}
\end{center}
\caption{{\small Azimuthally averaged radial
pressure (left axis - blue points) and entropy (right axis - red lines)
profiles for 3C\,449.
The uncertainties on pressure and entropy are
at $\sim$ 90\% confidence.
}}
\label{all_profiles}
\end{figure}

Using the temperature and surface brightness profile, we compute the
electron density profile.  We fit an analytic approximation to the best-fitting values
of $n_{\rm e}(r)$
$$
n_{\rm e} = 3.08 \times 10^{-3} \times \left[1 + \left(\frac{r}{15.41} \right)^2 \right]^{-\frac{3}{2} \times 0.29} {\rm cm}^{-3}
$$
along with $kT(r)$ to obtain profiles of
pressure and entropy, which are shown in Figure~\ref{all_profiles}.
The pressure profile shows a smooth decline as a function of distance from core;
whereas the entropy profile shows a gradual increase to 10~kpc, followed
by a rise to 80 kpc (an entropy bump) and finally
a flattening, which is consistent with the results of \citet{sun09}.

\subsection{Temperature Maps}

\begin{figure*}[ht]
\begin{center}
\begin{tabular}{ll}
\includegraphics[width=8.7cm]{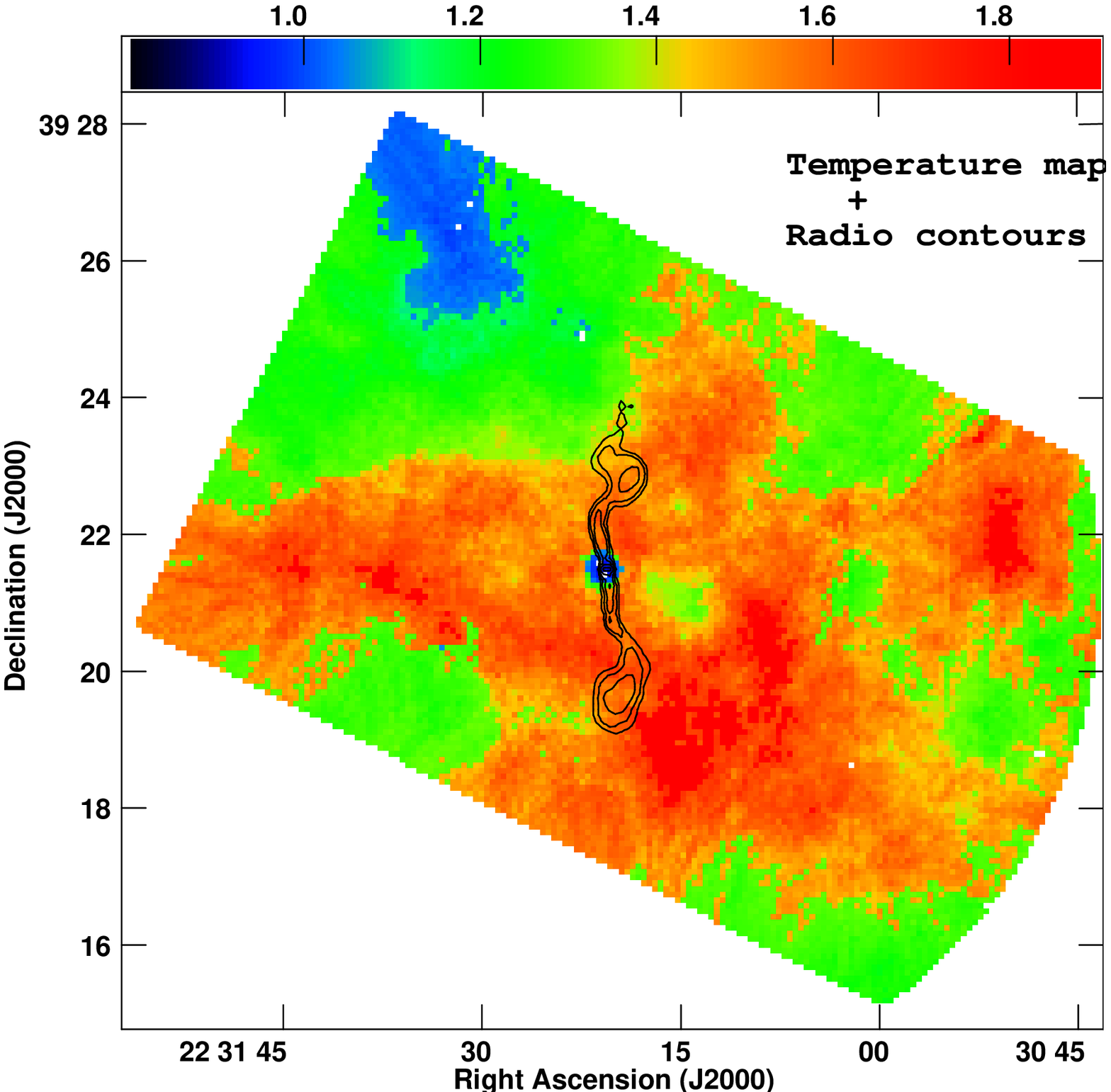} &
\includegraphics[width=8.7cm]{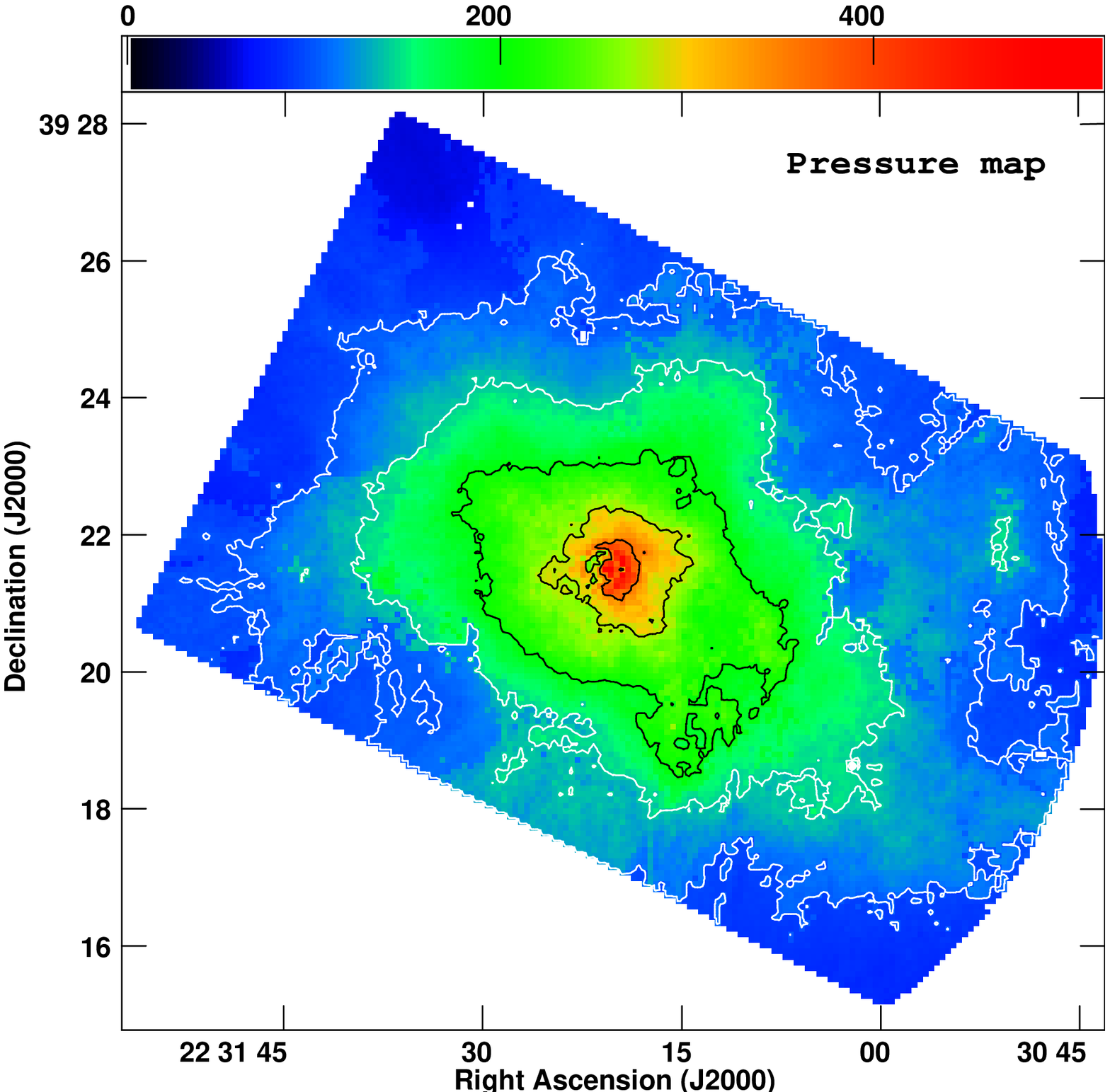} \\
\includegraphics[width=8.7cm]{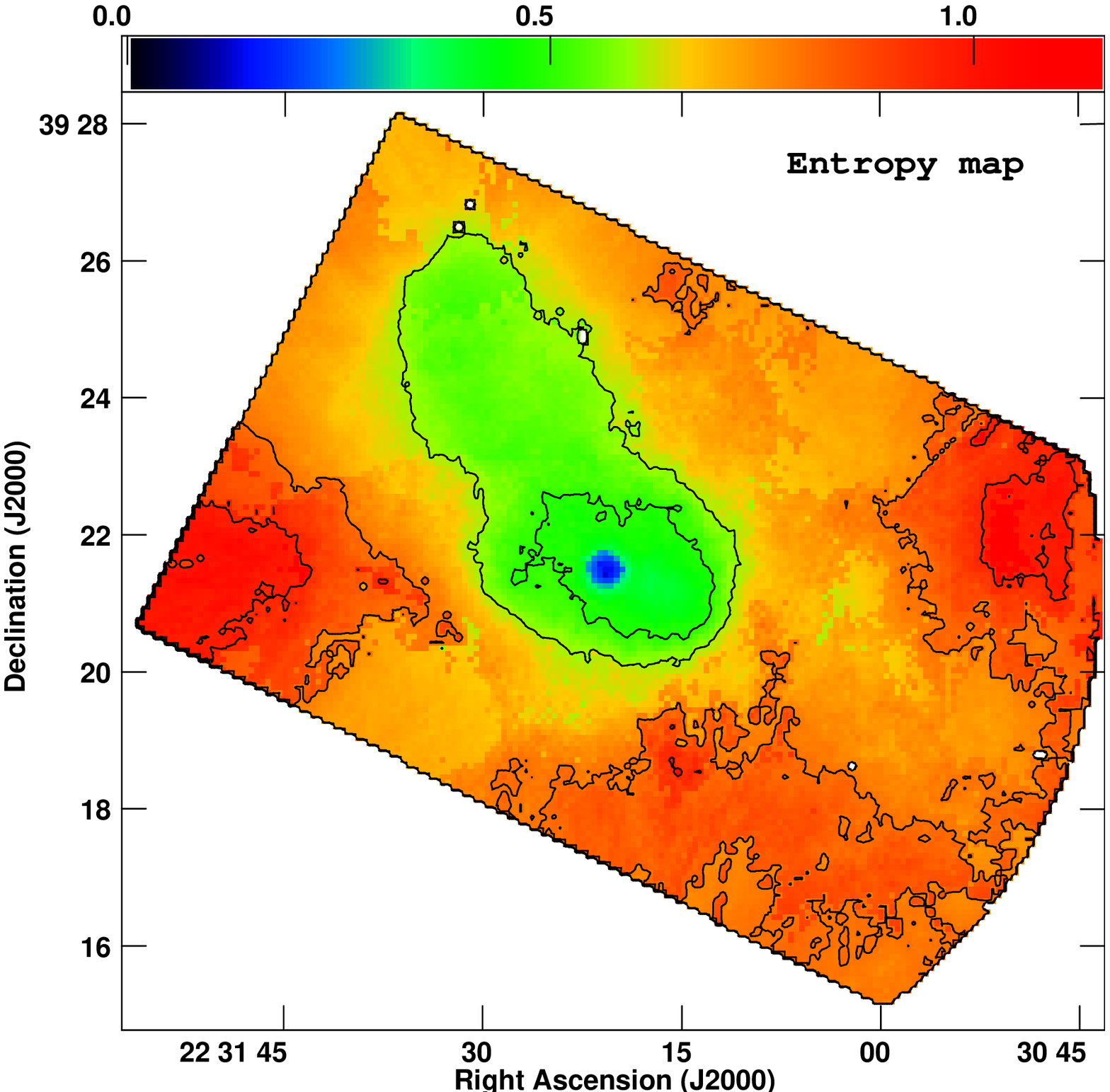}
\end{tabular}
\end{center}
\caption{{\small Thermodynamic maps of the 3C 449 group gas derived 
from the \textit{Chandra} data using the technique described in \citet{churazov03}.
Top left map has radio contours overlaid on the temperature map, top right
is the pseudo-pressure map, and bottom left is the pseudo-entropy map.
The color bar scale gives the temperature in keV in the temperature map and
is in arbitrary units in the pressure and entropy maps.}}
\label{tpae}
\end{figure*}

We created temperature maps
of the hot ICM.  These maps were created using the method developed
by \citet{ejos05} and \citet{maughan06}, which is outlined in
full by \citet{randall08}.
Briefly, after excluding the AGN and other point sources,
the 0.5--5.0~keV X-ray image was binned
in 5$^{\prime\prime}$ $\times$ 5$^{\prime\prime}$ pixels size and at
the position of each pixel, source and background spectra are extracted in a
circular region surrounding the pixel. 
The radius of this region at each pixel is defined
adaptively so that the extracted region contains 1000 background-subtracted
counts.  For each spectrum, a single {\sc VAPEC} model was fitted in the
energy range 0.5--5.0 keV, with the absorption column density set at the
Galactic value.  The abundance for O, Fe, and Si fixed to 0.12, 0.32 and 0.56
respectively, the abundance for the remaining elements fixed
to 0.5~$\times$ Solar, and the temperature allowed to vary.
The temperature map is presented in the top panels
of Figure~\ref{tpae}
and the typical uncertainties in the temperatures are $\sim$10\% at 90\% confidence.
Note that the formal statistical uncertainty in every pixel is roughly identical,
but the pixels are not independent, so any small scale structures are not significant.
The large scale (tens of arcsec)
structures presented in different colors are significant, however.

The temperature maps reveal a rich
diversity of patchy temperature structures, the most significant of which
are a small cool core (shown as the blue at the center of
the image) and the prominent cool $<$~1~keV filament region.
The latter is a wedge between P.A. 10$^\circ$
and 35$^\circ$, coincident with the asymmetry
seen in the X-ray brightness, which we call the ``cool filament''.

\begin{figure}
\begin{center}
\begin{tabular}{l}
\includegraphics[width=8.4cm]{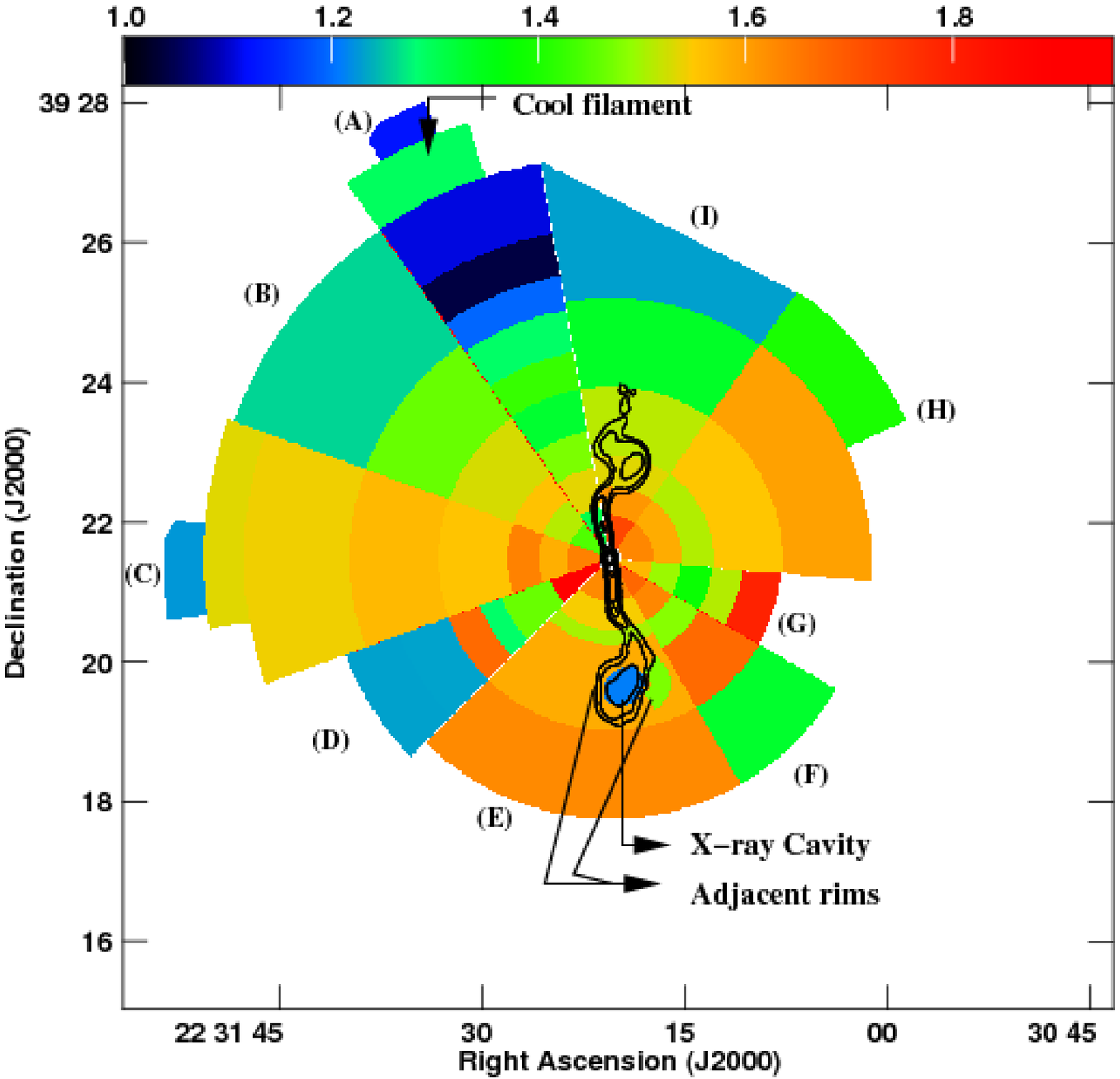}
\end{tabular}
\end{center}
\caption{{\small Projected temperature map using the merged data.
Point sources are excluded.
The color bar gives the temperature scale in keV.
The confidence ranges for regions with net counts $\gtrsim$ 3000 are low,
0.04--0.07~keV (inner sectors, close to the core),
whereas for regions with low net counts $\lesssim$ 1000 are high,
0.13--0.45~keV (outer sectors) at the 90\% confidence level.}}
\label{temp_region}
\end{figure}

We also generated projected pseudo-pressure and pseudo-entropy maps,
similar to those presented by \citet{churazov03} and \citet{randall08}, 
shown in the bottom left
and bottom right panels of Figure~\ref{tpae}, respectively.
The column density map $n_{\rm col}$ was estimated using
the square root of the 1.2-2.5~keV \textit{Chandra} image,
as in \citet{forman07}.
Combining this with the temperature map $kT$, we compute
the pseudo-pressure, $P$, and pseudo-entropy, $K$, at each pixel,
given by $P=n_{\rm col}\,kT_{\rm pix}$, and
$K=kT_{\rm pix}\,n_{\rm col}^{-2/3}$, respectively.
Unlike temperature maps, the symmetric pressure distribution
which peaks at the X-ray surface brightness peak indicates
the hydrostatic nature of 3C\,449.
The pressure map shows no hint of sharp changes at the
locations of the radio lobes.
The pressure map, which is near hydrostatic equilibrium,
is consistent with simulations of sloshing galaxy cluster cores, where the pressure distribution
stays the same during the sloshing \citep{zuhone10}.
The structure in pseudo-entropy is similar to what we see in the
temperature map;
i.e.  the pseudo-entropy map shows a small low-entropy core,
and a low-entropy tail to the northeast, which is
coincident with the cool filament/tail.

This method produces a temperature map with
correlated pixels \citep{ejos05, maughan06}, 
and the scale over which the pixels are correlated
is not clear from the temperature map alone. 
Therefore, as a check we produced a second map using a number
of independent spectral regions from the
background-subtracted, exposure-corrected \textit{Chandra} image
and residual image.
In particular, we considered several wedges based on the
\textit{Chandra} 0.5--2.0~keV surface brightness images 
(Figures~\ref{counts_image}, \ref{images}
and \ref{residual}) and the thermodynamic image (Figure~\ref{tpae}, top panel
images).
We further divided each wedge, named from (A) to (I) into several
elliptical annuli (1, 2, ..., n, where annulus ``1'' is close to the core
and ``n'' is farthest from the core) to investigate temperature structure.
Spectra for these annular quadrants were fitted with the {\sc VAPEC} model.
The abundance was again fixed to 0.12, 0.32, 0.56 $\times$ solar for
O, Fe and Si, respectively and 0.5 $\times$ solar for rest of the elements.
The confidence ranges for regions with net counts $\gtrsim$ 3000 are small,
0.04--0.07~keV (90\% confidence, $\chi^2$ 0.98--1.58 for 88--239 DOF),
whereas for regions with low net counts ($\lesssim$ 1000) the confidence ranges are large,
0.13--0.45~keV (90\% confidence, $\chi^2$ 0.58--1.62 for 18--239 DOF).
A crude temperature map (in which all the bins are
independent) along with our spectral extraction
regions are shown in Figure~\ref{temp_region}.
This crude temperature map with independent pixels confirms the general features
seen in the adaptive map including the cool filament to the northeast and the lack
of any large temperature jumps associated with other features in the gas.

\subsection{Surface Brightness Edges}
\label{large_scale_diffuse}

We find two edges in the surface brightness distribution,
one 97.9$^{\prime\prime}$ (= 32.6~kpc) southeast of the nucleus and
another 96.7$^{\prime\prime}$ (= 32.2~kpc) to the west, as marked in Figure~\ref{images}.
The presence of these edges in the surface brightness distribution implies a sharp change in the gas density or temperature
(or both) of the gas across the edges.  \textit{Chandra} has observed a large number of similar features
in other clusters, such as Abell\,1795 \citep{maxim01}, Abell\,3667 \citep{vikhlinin01a},
M\,87 \citep{forman2005,forman07}, Hydra\,A \citep{nulsen2005}, MS\,0735.6$+$7421 \citep{mcnamara05},
Abell\,1201 \citep{owers09} and 3C\,288 \citep{Laletal2010},
and they are generally
attributed to three phenomena:  merger cold fronts, sloshing
of cluster cores, and shocks due to nuclear outbursts \citep{maxim07}.
To investigate whether the observed surface brightness
decrease is associated with shocks or cold fronts, we extracted surface
brightness profiles and spectra in several regions of interest.
Following the analysis of \citet{maxim07},
we determined the temperature and pressure across the edge to evaluate
which of the three scenarios; merger cold-front, sloshing cold-front, or supersonic inflation
of radio lobes, is most plausible.

\begin{figure*}[ht]
\begin{center}
\begin{tabular}{ll}
\includegraphics[width=5.5cm,angle=-90]{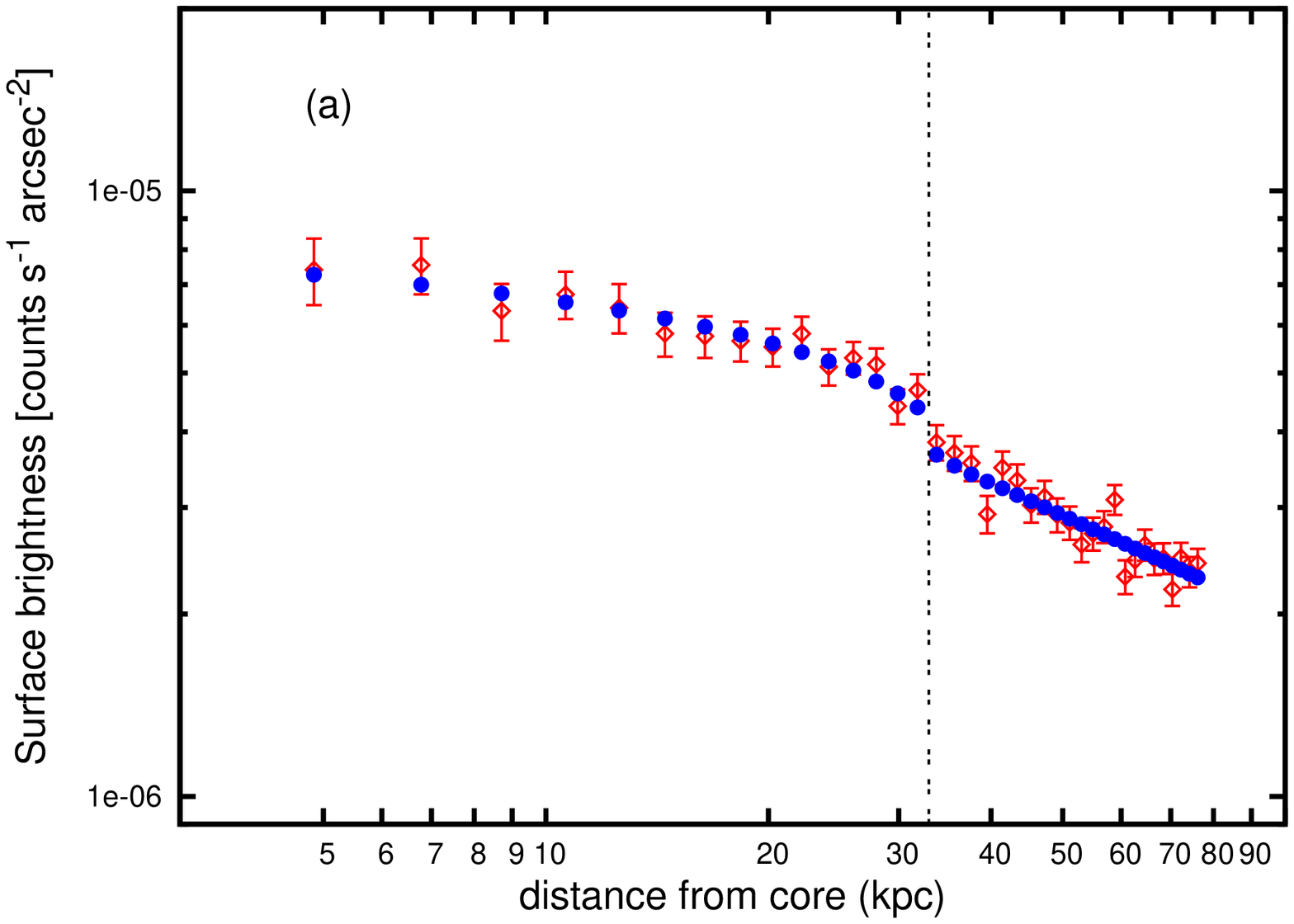} &
\includegraphics[width=5.5cm,angle=-90]{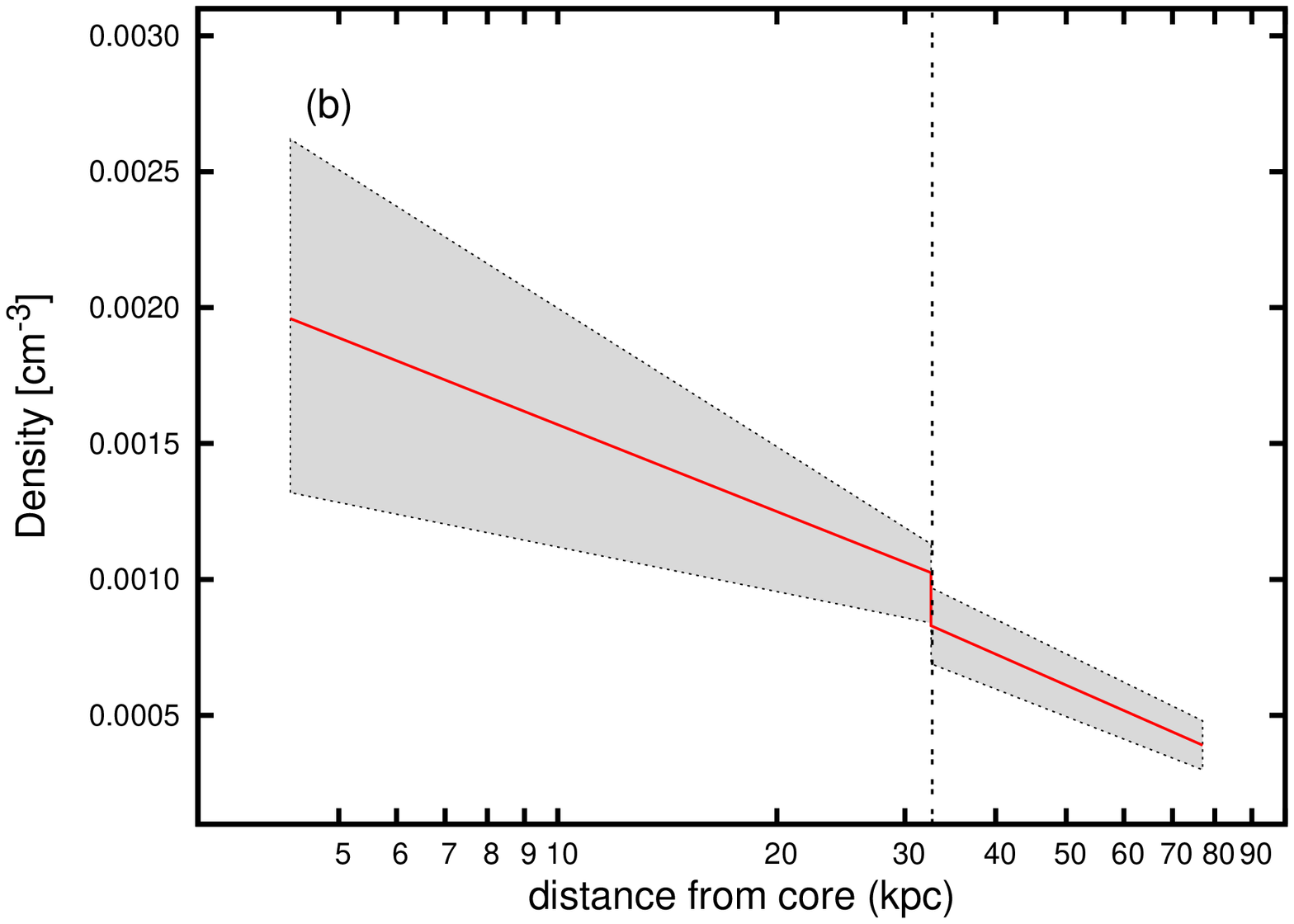} \\
\includegraphics[width=5.5cm,angle=-90]{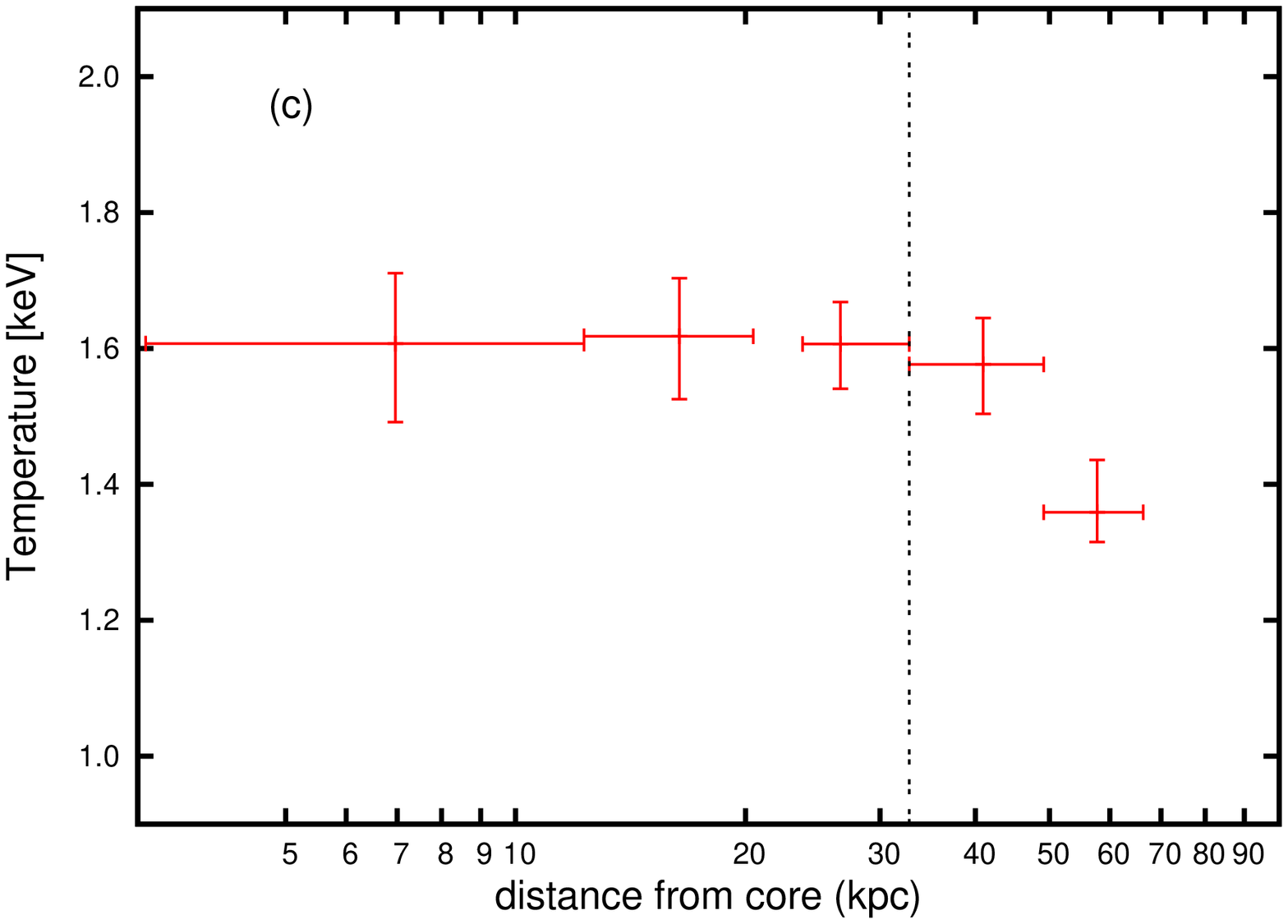} &
\includegraphics[width=5.5cm,angle=-90]{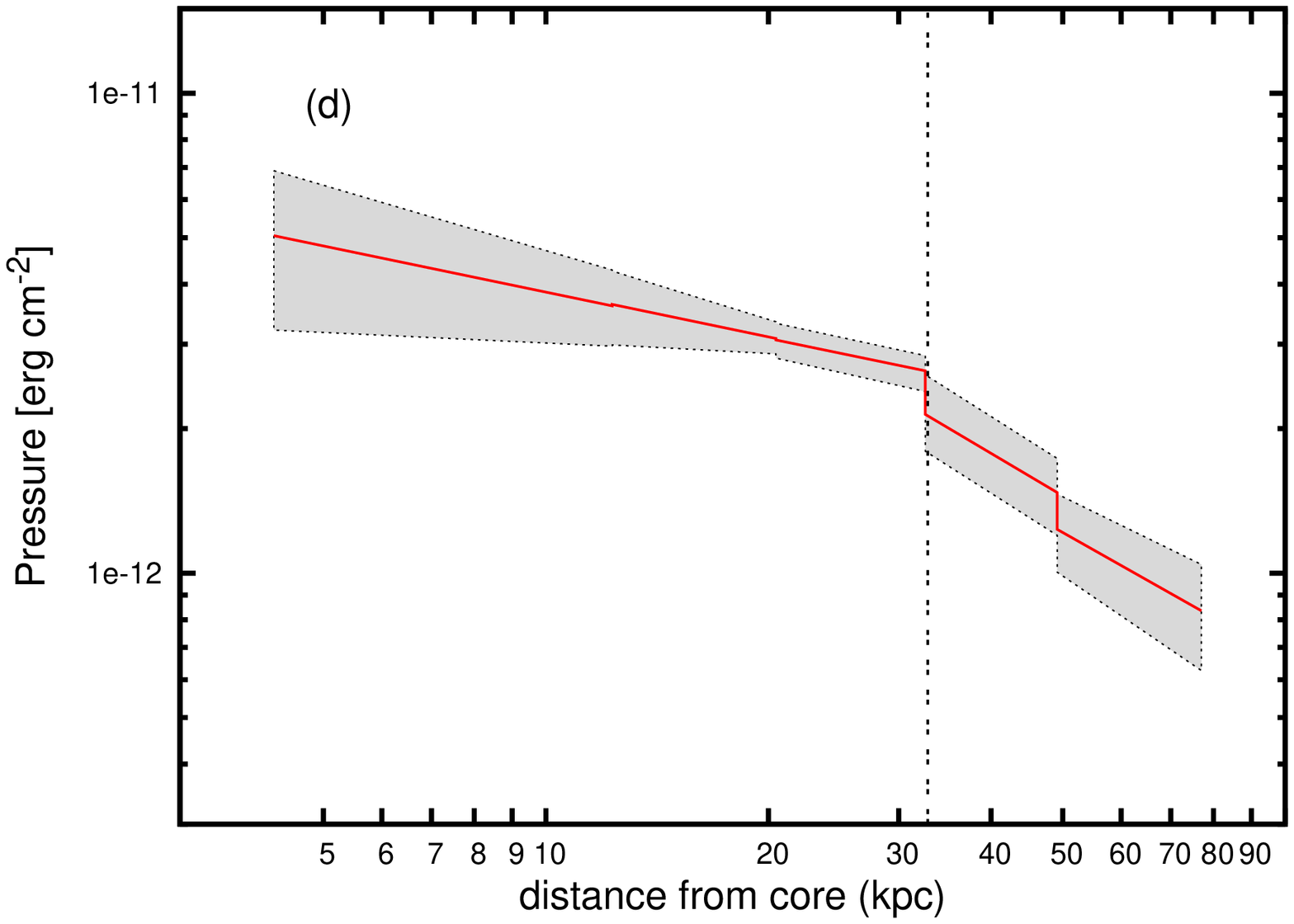}
\end{tabular}
\end{center}
\caption{{\small Surface brightness edge in 3C\,449 showing the
sector between P.A. = 70$^\circ$ and P.A. = 160$^\circ$ (wedge to the east) in the
0.5--2~keV band.
Panel (a) shows eastern X-ray surface brightness profile.  The filled
circles are the model surface brightness values that correspond to the best-fit gas
density model shown in panel (b), and the  temperature profile in panel (c).
Panel (d) shows pressure profile obtained from the temperature and the density profiles.
The errors are at 90\% confidence level and the
vertical dashed line in each panel shows the position of the density jump.}}
\label{se_sb_profile}
\end{figure*}
\begin{figure*}[ht]
\begin{center}
\begin{tabular}{ll}
\includegraphics[width=5.5cm,angle=-90]{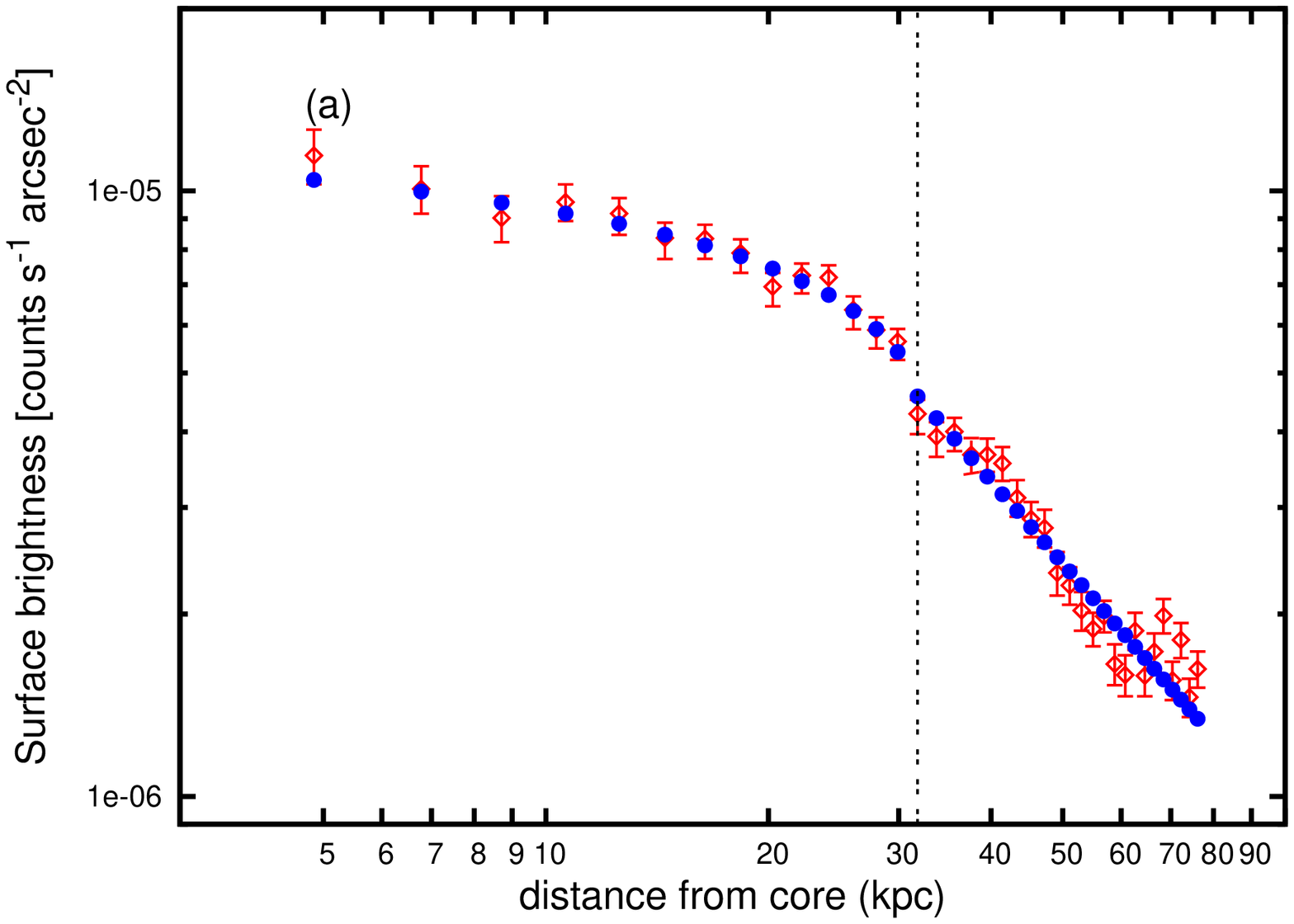} &
\includegraphics[width=5.5cm,angle=-90]{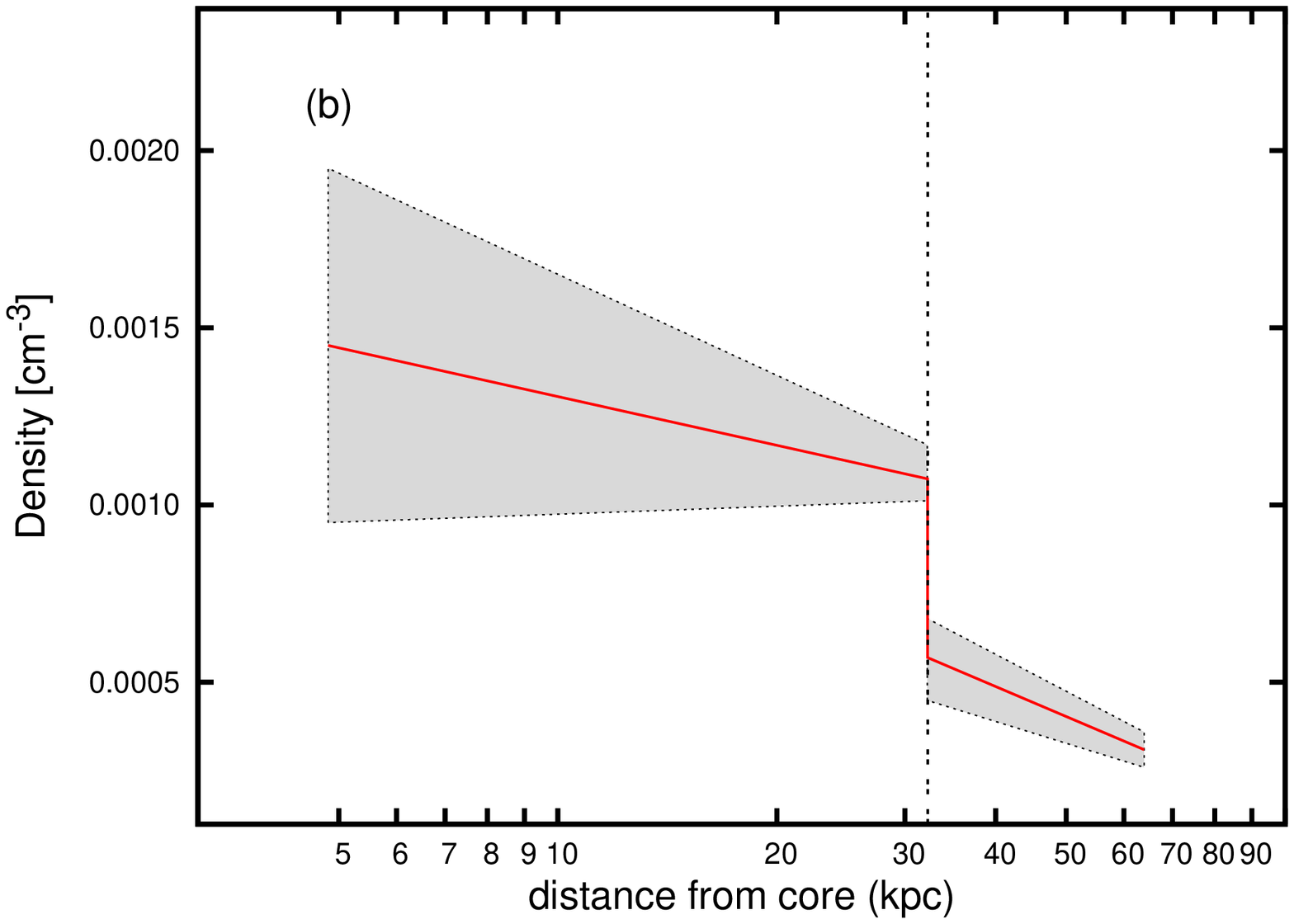} \\
\includegraphics[width=5.5cm,angle=-90]{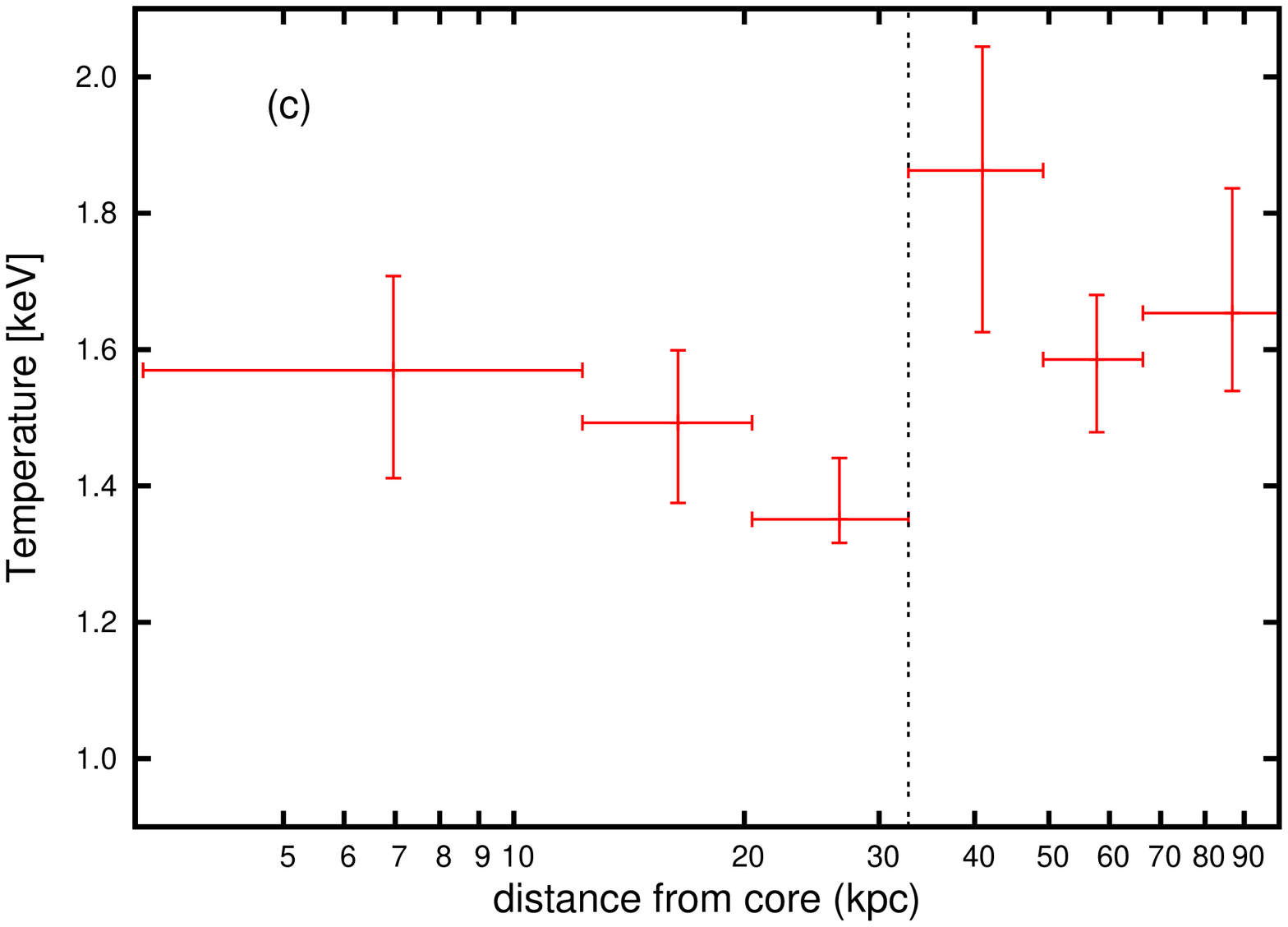} &
\includegraphics[width=5.5cm,angle=-90]{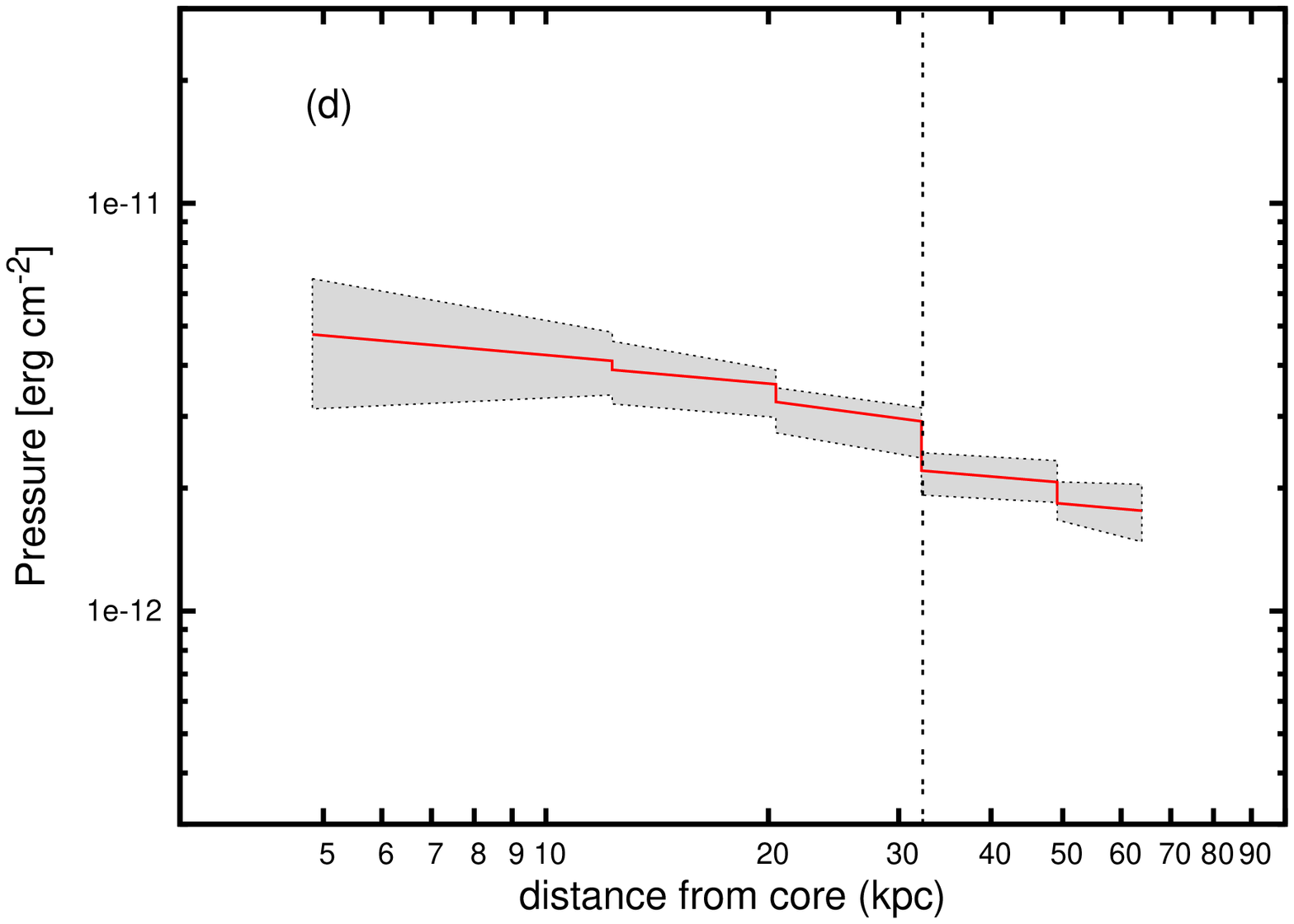}
\end{tabular}
\end{center}
\caption{{\small Surface brightness edge in 3C\,449 showing the
sector between P.A. = 230$^\circ$ and P.A. = 270$^\circ$ (wedge to the west) in the
0.5--2.5~keV band.
Panel (a) shows X-ray surface brightness profile.  The filled
circles are the model surface brightness values that correspond to the best-fit gas
density model shown in panel (b), and the temperature profile in panel (c).
Panel (d) shows pressure profile obtained from the temperature the density profiles.
The errors are at 90\% confidence level and the
vertical dashed line in each panel shows the position of the density jump.}}
\label{w_sb_profile}
\end{figure*}

We fitted {\sc VAPEC} models to annuli in two sectors (marked in
Figure~\ref{images}) centered on 3C\,449.
The vertex of the annuli was fixed at the nucleus,
but their radii were adjusted to so that three annuli were interior and
two annuli were exterior to the edge.
The goal of this spectral fitting was to determine whether the gas temperature
interior to the edges was hotter or cooler than the exterior gas temperature.
Only the temperatures and normalizations were free parameters in these fits.
Plots of the surface brightness and temperature profiles for the two
different sectors
as a function of radius from the active nucleus
are shown in panels (a) and (c) of Figures~\ref{se_sb_profile} and~\ref{w_sb_profile}, respectively.
For the southeastern wedge, we measured temperatures of
1.61$^{+0.06}_{-0.07}$~keV inside the jump and 1.51$^{+0.07}_{-0.06}$~keV
outside the jump.  For the western wedge, we measured temperatures of
1.35$^{+0.09}_{-0.03}$~keV inside the jump and 1.86$^{+0.18}_{-0.23}$~keV
outside the jump.
The temperature ratios for the southeastern and western edges
are 1.02$^{+0.06}_{-0.07}$ and 0.72$^{+0.08}_{-0.09}$, respectively.  
The uncertainties are at 90\% confidence.

We fitted the surface brightness across the edge in the
sectors shown in Figure~\ref{images} in the energy range 0.5--2.0 keV with
a broken power-law density model to determine the density and pressure jumps across the
edges.  To minimize projection effects, ideally one would prefer to measure
the deprojected temperature profile across the edges; however,
the limited number of photons and small temperature differences do not permit this 
and we use the projected
temperatures measured above.
The lack of large temperature variation suggests that projection effects
are not large.
The deprojected density and pressure profiles
as a function of distance from the cluster center are shown in
Figures~\ref{se_sb_profile} and~\ref{w_sb_profile},
panels (b) and (d), respectively.
Both brightness profiles have a characteristic shape corresponding to
an apparent abrupt jump in the gas density.
Best-fit radial density models consisting of two power-laws with a jump
are shown in panel (b),
and the projected model surface brightness profiles are overlaid on the data in panel (a) of
Figures~\ref{se_sb_profile} and~\ref{w_sb_profile}.
From the amplitude of the best fitting surface brightness model,
we derived a density jump of 1.23$^{+0.20}_{-0.18}$ for the
southeastern edge and 1.89$^{+0.18}_{-0.21}$ for the western
edge.  The confidence ranges for the density jumps were
computed from the extremes of the 90\% confidence ranges for
the best fitting surface brightness model.
We assumed spherical symmetry in the deprojection of the surface brightness
to derive the densities.

The X-ray morphology of the filaments around the inner southern
radio lobe is similar to features seen in nearby radio galaxies.
These structures are formed when AGN jets push into the local
ICM, evacuating cavities, and often creating bright rims of X-ray
emission from the displaced gas.  In more evolved remnant
cavities, the rims tend to be cooler and more dense than the
nearby ambient ICM, as in Abell 2052 \citep{blanton03} and
Perseus \citep{fabian2006}, whereas in more recent outbursts
they often show higher temperatures associated with shocks,
as in NGC 4552 \citep{machacek06}, Hercules A \citep{nulsen2005},
Centaurus A \citep{croston09}, and
NGC\,5813 \citep{randall11}.
To test for a temperature difference in the rims, we extracted spectra
from the southern bubble and a region on the rim just
outside the bubble, subtracted a local background
region, and fit each with an absorbed VAPEC model with the abundance fixed
(see Section 3.1).
We find best-fitting temperatures of
$kT_{\rm E-rim}$ = 1.59$^{+0.41}_{-0.25}$ for the east rim,
$kT_{\rm W-rim}$ = 1.48$^{+0.32}_{-0.22}$ for the west rim and
$kT_{\rm cavity}$ = 1.21$^{+0.15}_{-0.16}$ for the cavity,
inside the rim.
The (projected) temperature of the gas in the rims is consistent within
uncertainties with that of the ambient gas.
We conclude that these rims are most likely low entropy 
gas that has been uplifted by the radio bubbles and not shocks.

Finally, there is a clear decrease in X-ray surface brightness at the position
of the southern inner jet.  We labeled this as the 'tunnel' in Figure~\ref{images}.
To test the significance of the tunnel-like feature, we extracted
the total count rate in evenly spaced bins across the radio jet
shown in Figure~\ref{images} (left panel).  Figure~\ref{tunnel} shows two distinct peaks
adjacent to a clearly visible valley above the nominal surface brightness.
If we take an average count rate from the highest three bins in each
peak and compare it with the average count rate from the lowest
three bins in the valley, we find that the gap is significant at about
the 1.9$\sigma$ level.
It seems that the gas may be compressed or the gas may have been uplifted
to create a ``sheath-like" feature,
and the tunnel is filled with the relativistic plasma.
Assuming the radius of the radio jet to be equal to the width of the tunnel
as seen on the sky, we estimate the decrement in surface brightness due to the evacuated
tunnel.
That is, we assumed that the volume occupied by the jet is devoid of gas and that the jet
is in the plane of the sky.  We integrated the square of the density profile at the position of the jet
to compute the surface brightness decrement created by the jet. 
The predicted decrement in emission measure is $\sim$0.86,
which is consistent with the measured
decrement in emission measure of 0.75 $\pm$0.09 within 90\% confidence ranges.
This also suggests that the southern jet lies at or near the plane of the sky (i.e. the
decrement would be smaller if the jet were far from the plane),
which is consistent with the generally symmetric appearance of the inner jets.

\section{Discussion}

We report the detection of several previously unknown features in the hot gas
atmosphere around 3C 449.  
Both the surface brightness and temperature distributions
of the gas display a complex, asymmetric morphology within $\sim$100 kpc of the radio/X-ray core. 
Some features arise from the AGN activity, 
but others are better explained by an earlier group merger.

\begin{figure}
\begin{center}
\begin{tabular}{l}
\includegraphics[width=6.0cm,angle=-90]{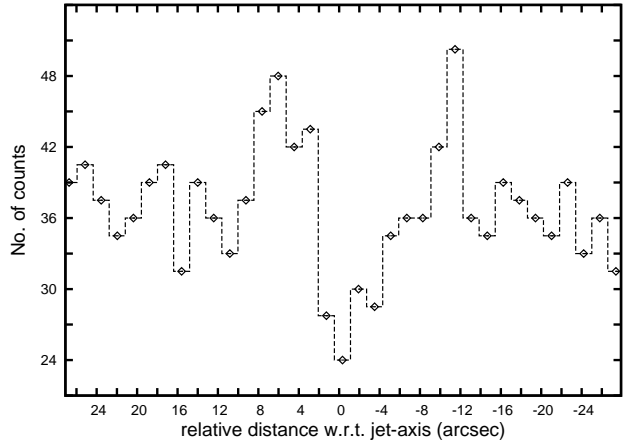}
\end{tabular}
\end{center}
\caption{Histogram showing distribution of the counts in the
energy range 0.5--2.0~keV in evenly spaced binned regions taken
across the ``dashed" tunnel-like feature (shown in Figure 2) in the shape of a
box (56.6$^{\prime\prime}$ $\times$ 29.5$^{\prime\prime}$ and centered at
$\alpha$ = 22:31:20.289, $\delta$ = 39:20:58.51),
which is $\sim$31$^{\prime\prime}$ south of the nucleus.}
\label{tunnel}
\end{figure}

The first of the AGN-related features is a cavity at the position of the southern radio lobe
which is surrounded by a rim of enhanced X-ray emission. 
The temperature of the rim is marginally higher than in the cavity,
but is consistent with that of the ambient gas within the large error bars.
Furthermore, there is a tunnel-like feature in the X-ray emission 
connecting the southern cavity and the group core (see Figure~\ref{tunnel}),  
which coincides with the southern radio jet.
The radio jet seems to have created the tunnel,
either by compressing the gas or by uplifting the gas
to form a sheath-like feature.
If the former is true,  the compressed gas should have been hotter
than the surrounding gas and very short-lived since it would
be over-pressurized, whereas in the latter case the
uplifted gas should have been cooler than the surrounding gas
and in rough pressure equilibrium.  It is unlikely that the jet is greatly
over-pressurized relative to the ambient gas, so the filaments surrounding the
tunnel are likely to be low entropy gas that has been uplifted by the jet.

There are two arc-shaped surface brightness discontinuities in the gas to the southeast
and to the west of the nucleus, at approximately $\sim$33.4 kpc from the group core. 
Surface brightness edges as seen in the gas of 3C 449
are commonly attributed to three phenomena:  
remnant cold-core merger fronts, sloshing fronts, or supersonic inflation of radio lobes.
In the presence of the obvious AGN activity in this group one might attribute 
these brightness edges to an AGN-driven shock,
but there are several reasons why this is probably not the case.
First, we find no evidence for a temperature increase of the gas interior to the
edges as would be expected for a shock (see Figure~10 of \citet{forman07}). 
On the contrary, the gas interior of the western edge seems 
cooler than the exterior gas, typical of a cold front, and there is little
temperature jump across the eastern edge. 
Even though the constraints on the temperature are not sufficient  
to rule out a weak ($M<$1.5) shock, there is no evidence for a 
significant pressure jump either, which again favors an interpretation as cold fronts. 
Second, the AGN-driven shock is expected to enclose the cavities and radio lobes, 
whereas in 3C 449 the southern inner lobe is outside the discontinuity
assuming even a small ($\sim$15$^\circ$) angle of the jet to the plane of the sky.
It is hard to see how the lobes could be driving a shock if they are outside the shock.
Third, the radio morphology doesn't support the shock interpretation.
Instead of appearing as a momentum dominated FR II type source like
Cyg A \citep{wilson06} or an over-pressurized energy driven
bubble like Cen A \citep{kraft03}, both inner radio lobes are bent and wispy, suggesting a more gentle, nearly
quasi-static interaction with the ambient gas.
Fourth, the distance between the edge and the nucleus is larger in the
E/W axis than the N/S axis where the lobes are located.
Given that the radius of the lobes is only slightly larger than the jet, the shocks
should either surround the lobe if the lobe is inflating supersonically or should form
an oval around the jet/lobe pair if the jet is momentum dominated.
If the edge is in fact a density/temperature discontinuity created by an AGN-driven shock,
it is hard to see how the shock could propagate faster perpendicular to the 
axis of the driving piston than parallel to this axis.
There are thus multiple lines of evidence to suggest that this edge is not an AGN-driven shock. 

Even if the edges are not AGN-driven shocks,
there does, however, appear to be a clear connection between the 
the gas structure and the AGN because both inner jets bend eastward and expand to
form the inner radio bubbles at the same group-centric radius 
where the edges are found. 
We discuss this connection below.

\subsection{Merger Dynamics}

\subsubsection{Remnant merger core}

We argued that the brightness edges in 3C 449 
are unlikely to be an AGN driven shock, but more closely resemble cold fronts. 
A remnant or merger cold front is the interface between the dense cool gas core 
of a subcluster or subgroup and the hotter ambient gas the subgroup is 
moving through. Here, the cold fronts surround the group core itself and 
could imply a motion of the group core moving through the ambient group gas. 
Approximating the cool gas core as a blunt body, 
\citep{vikhlinin01a,ll59},
the pressure ratio between the pressure at the stagnation point and 
in the free stream region depends on the Mach number of the moving body.
The conservative confidence intervals of pressure ratio are
1.25$^{+0.20}_{-0.23}$ and 1.35$^{+0.23}_{-0.21}$ for the
southeast and west surface brightness edges, respectively.
These correspond to a subsonic motion with  Mach numbers, $M$, for the southeast and west edge of
0.53$^{+0.20}_{-0.23}$ and 0.62$^{+0.18}_{-0.17}$
(Figure 6 of \citet{vikhlinin01a}) .
The implied velocity in physical units
are 354.3$^{+130.3}_{-150.2}$ km~s$^{-1}$ and
414.8$^{+119.6}_{-113.0}$ km~s$^{-1}$, respectively, 
where we used the gas temperature
$kT$ = 1.66$^{+0.04}_{-0.05}$~keV, $\gamma$ = 5/3
and the mean molecular weight of the intracluster plasma $\mu$ = 0.6.
The presence of two edges roughly 180$^\circ$ apart strongly argues against these features
being the result of a remnant core.  In addition, the relatively low (i.e. subsonic) velocity
is not consistent with a remnant merger core at the center of the group potential.  A remnant
core that has infallen to the center of the group should be supersonic.
Thus, we reject the remnant core scenario.

\subsubsection{Sloshing}

Sloshing cold fronts arise when an initially close to 
hydrostatic cluster or group atmosphere is slightly 
offset from its equilibrium position, e.g.~by a minor or 
distant merger. The offset gas oscillates inside the central 
potential well, which leads to the formation of arc-like cold fronts 
wrapped around the cluster core (e.g.~\citealt{yago06,RoedigerA,zuhone10}). 
All but straight head-on mergers transfer angular momentum, and the 
sloshing takes on  a spiral-like appearance in the orbital plane. If the 
line-of-sight is perpendicular to the orbital plane, also the cold fronts form a 
typical spiral structure. For line-of-sights parallel to the orbital plane, the cold 
fronts are arranged as  staggered arcs on opposite sides of the cluster.

The arrangement of the cold fronts in the 3C 449 group neither matches perfectly the cold remnant 
nor the pure sloshing scenario. Remnants typically have a surface brightness discontinuity on only
one side, not two.  At the group center, a remnant core should be moving
supersonically with a clear pressure discontinuity indicative of its infall
velocity and perhaps a bow shock (although the shocks are often not visible due to
projection or limited S/N of the data).  Sloshing cold fronts on opposite sides of the core usually appear at 
different cluster-centric distances. If the southeastern and western cold front 
form a continuous sloshing front around the south of the group core, 
we expect another front at a different distance to the north, but there is 
no indication for such a brightness discontinuity in the data. The observed 
cold filament is explained by neither scenario. 

\begin{figure*}[ht]
\begin{center}
\begin{tabular}{ll}
\includegraphics[width=8.5cm]{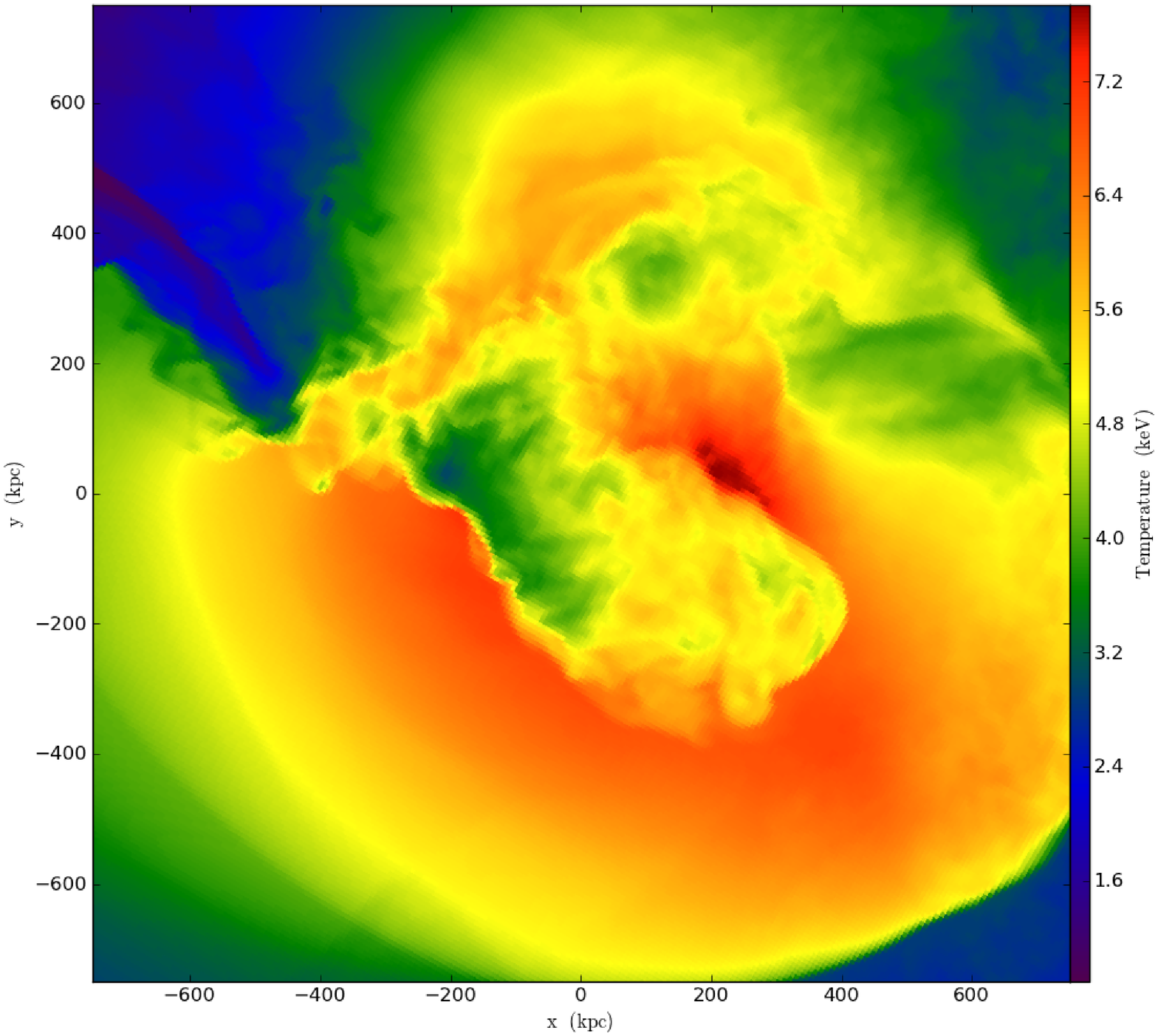} &
\includegraphics[width=8.5cm]{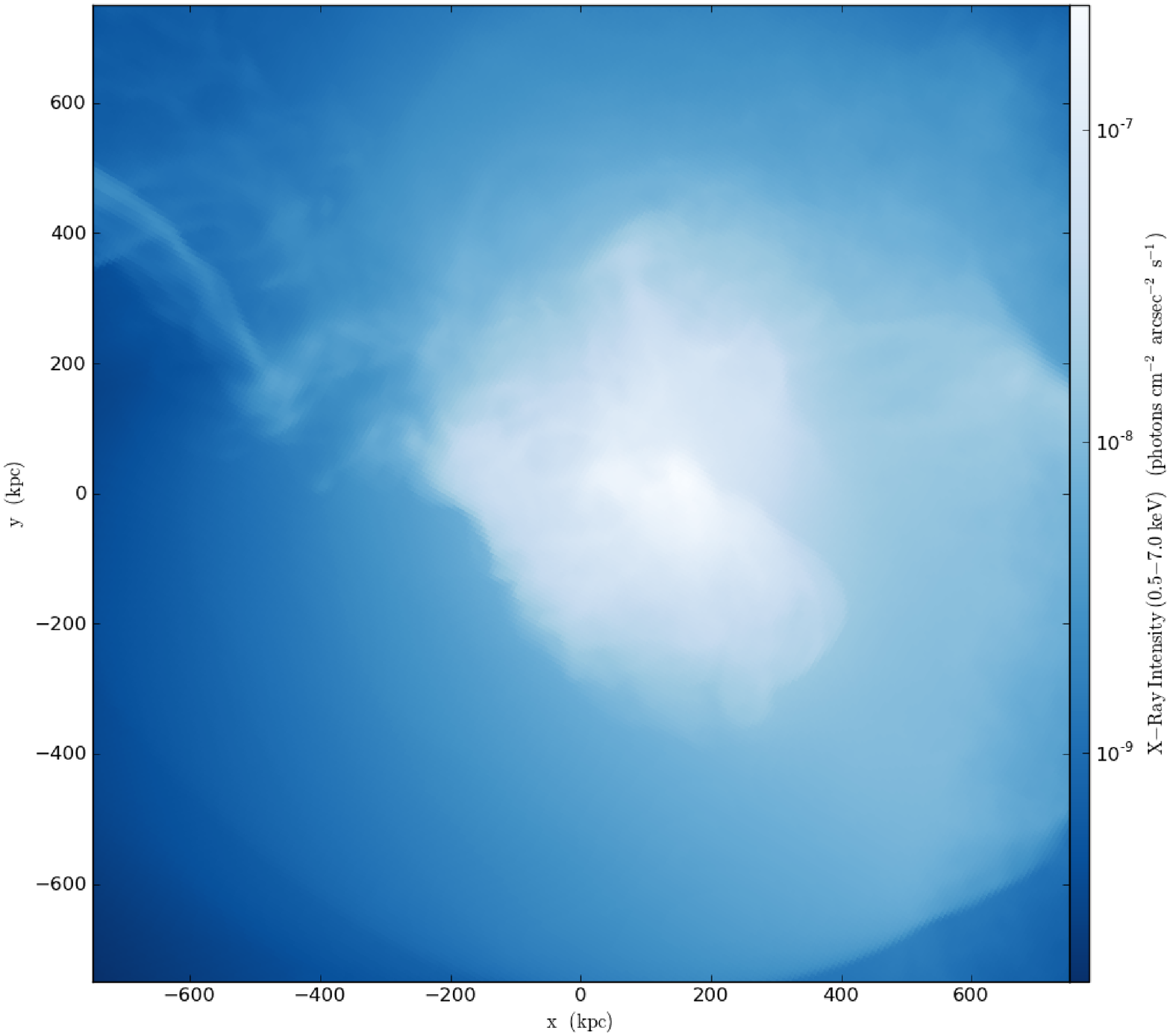}
\end{tabular}
\end{center}
\caption{{\small Slice through the temperature (left panel) and
projected surface brightness (right panel) from our cluster merger simulation.
Each axis is in kpc.  Major tick marks
indicate 200 kpc distances. The color scheme in the left panel temperature image is in
shown on the cold bar on the right side of the image.
The cold filament and asymmetric core emission are clearly seen
and some of the features are highlighted.
Note that these images come from a simulation on a galaxy cluster scale
(and so the linear scales are larger and temperatures higher)
and we use this good match for the galaxy group with an
appropriate scaling.  The key point is that the cool filament is a standard feature
of late stage cluster mergers.  Note that the sloshing cold front can be seen at the bottom
right of both images.}}
\label{sim_imgs} 
\end{figure*}

However, all these features can be explained if the 3C 449 group 
is in a late stage of a major merger. Figure~\ref{sim_imgs} 
displays a mock X-ray surface brightness and temperature map of a binary cluster merger 
with a mass ratio of 1:3 and with
a large impact parameter \citep[see Figure~8, t=4.0 Gyr epoch of][for entropy map of this
simulation]{zuhone11}. The smaller subcluster or subgroup has had its
entire gaseous atmosphere ram pressure stripped.  This stripped gas forms a long 
cool tail or filament pointing away from the major group center. 
Due to the large impact parameter, the core of the larger group is 
not destroyed, but indeed sloshing motions are triggered. For low 
mass ratios and also at late stages the sloshing is not as well-ordered 
as in a minor merger, but more irregular, and the spiral or staggered 
arc pattern is distorted or missing.  Scaling timescales and
spatial scales of the cluster merger in Figure~\ref{sim_imgs} 
to a group merger appropriate for the 3C 449 group, we estimate that the core 
passage occurred $\sim$1.3--1.6 Gyr ago.

We emphasize that in this interpretation the surface brightness discontinuities we see in our 
X-ray image are the secondary, inner,
sloshing cold fronts, not the primary ones typically seen in X-ray images.
We note that in the residual surface brightness image in
Figure~\ref{residual}, there is an excess of emission $\sim$115 kpc to the
southwest of the nucleus.  This excess is at roughly the position/distance of the primary
sloshing cold fronts (assuming that the surface brightness edges we see are in fact
the secondary sloshing cold front), and may denote the position of the primary front or
some larger scale asymmetry.  There is no evidence of the primary sloshing
edge in our data if we extend the profiles out to this distance, 
so whether it actually marks the position of the primary cold front
is unclear.  Sloshing can also introduce asymmetric features in the gas on large scales
since the core (both gas and dark matter) of the group is offset from the center
of the large scale gravitational potential.
Finally we mention that the inflation of the large scale radio lobes has also input significant
thermal energy into the gas and may well have distorted the appearance of the
primary sloshing cold front.  This interpretation could be evaluated with a deep Chandra
mosaic.

The jet axis in 3C 449 is close to the plane of the sky. If the orbital 
plane of the merger was also close to the plane of the sky, it would have 
introduced bulk flows in this plane. However, the outer radio jets are very 
straight out beyond 100 kpc, which may imply a significant inclination 
between the merger plane and the plane of the sky. 

\subsection{Radio Galaxy Dynamics}

Hydrodynamic simulations of jets crossing density discontinuities
with shear flow show that well collimated jets can become partially disrupted
\citep{loken95}, and the disruption of the jets of 3C 449 by this type of
edge was first predicted by \citet{katzstone}.
Assuming a moderate angle between the jets and the plane of the sky,
it is plausible that the interface between the inner jets and inner lobes occurs exactly at the
surface brightness edges seen in our data.
If this is the case, the jets must be inclined $\sim$15$^\circ$ relative to the
plane of the sky.
This angle is consistent with the angle of the radio-jet to the line-of-sight,
$\sim$75$^\circ$ estimated via relativistic motion of the inner (100 pc) jet
\citep{feretti99}.
Interestingly, one striking aspect of this radio source is the mirror symmetry
of the jets and outer lobes about an axis through the central nucleus.
The fact that both of the inner lobes of 3C 449 are bent
in the same direction to the west suggests that the gas exterior to
the edge is pushing them in that direction.
This is entirely consistent with the sloshing picture described above.
Since the gas is moving with the central galaxy and gas interior to the
sloshing cold front,
the jets interior to the edge remain straight and well collimated.
Once the jet crosses the edge, however, it is partially disrupted, entrains
group gas, and inflates to form inner radio lobes.
Additionally, the simulations of \citep{loken95} also show that this
disruption can drag cold gas from the interior out beyond the edge,
and we associate the filaments around the southern inner bubble with this
process.

The displaced gas from the tunnel at the position of the southern
inner jet has clearly
been deposited around it to create a sheath-like feature,
and the tunnel is filled with the radio-jet plasma.
To our knowledge this is the first case where we detect
the X-ray sheath enveloping an established and stable radio plasma jet.
The X-ray enhancements around the tunnel must be low entropy gas that
has been entrained and uplifted by the jet.  If it were shock-heated gas,
it would expand rapidly into the ISM and would be invisible.
The jet may be entraining material and developing an internal sheath in addition to
creating the observed cocoon of X-ray emitting gas. 
This internal sheath may be the steeper spectral index component surrounding the jet
reported by \citet{katzstone}.
The entrainment of a small amount of thermal gas may therefore have a dramatic effect
on the relativistic electron population and/or on the efficiency of particle
acceleration.

We can estimate the work done on the X-ray emitting gas by the inflation
of the inner bubbles. 
Assuming a spherical geometry for the southern bubble
(with radius of $\simeq$13~kpc, located at a distance
from the core of 1.9$^{\prime}$ $\simeq$ 39~kpc), we find an average
electron density of 1.1$\times$10$^{-3}$ cm$^{-3}$ and
pressure of 5.4$\times$ 10$^{-11}$ erg cm$^{-3}$.
The work required to inflate the cavity against this pressure
is $pV$ $\simeq$ 1.46 $\times$ 10$^{57}$ erg,
where $V$ is the volume of the cavity.
Including the internal energy of the cavity,
the total energy required to create both cavities, i.e.,
its enthalpy, would be 4$pV$ if it is filled with relativistic gas,
amounting to about $1.2\times 10^{58}$ ergs.
Furthermore, the age of the cavity can be estimated as the
time required for the cavity to rise the projected distance from
the radio core to its present location at the speed of sound, or
buoyantly at the terminal velocity \citep{birzan04},
$\sim$7 $\times$ 10$^{7}$ yr.
The mechanical power of the current outburst inflating the inner lobes is,
L$_{\rm mech}$ = $2 \times 4 pV$/age$_{\rm cavity}$, a relatively modest
$\sim$5.7 $\times$ 10$^{42}$ erg~s$^{-1}$.
We note that this only gives a lower limit to the jet power since the flow of
plasma may continue into the outer lobes.

Finally we mention that the unusual entropy profile of the group gas cannot be
explained by the supersonic inflation of the inner radio lobes.  
These inner lobes do not appear to be inflating supersonically, the
entropy peak lies beyond the position of the inner lobes, there is
clear evidence that this group is in the late stages of a merger, and the larger scale
lobes are also probably evolving buoyantly.  Hydrodynamic simulations of
sloshing show transient oscillations of the entropy profiles \citep{RoedigerA}.
We therefore attribute this unusual structure in the entropy profile to merging,
not nuclear activity.  Interestingly, the other group studied by \citet{sun09}
that has an unusual entropy peak is UGC\,2755.  This group also hosts a powerful
radio galaxy, and examination of the archival Chandra data on this object
also shows features in the X-ray gas on scales larger than the radio outburst
indicative of sloshing/merging.

\section{Conclusions}

In this paper we presented an analysis of the hot gas around the canonical FR\,I radio
galaxy 3C\,449 based on 140 ks \textit{Chandra} X-ray
data and archival radio data.
The combined \textit{Chandra} images reveals not only
complex, asymmetric morphologies of surface brightness and temperature
but an asymmetric core emission, sharp edges,
a ``tunnel-like'' feature and a cavity associated with the southern inner jet and lobe, respectively.
An earlier, shorter \textit{Chandra} 30~ks observation of the group gas showed
an unusual entropy distribution and evidence for a surface brightness edge in the gas.
We find in our deeper dataset that this edge is probably a sloshing cold front due to
merger $\lesssim$1.3--1.6~Gyr ago.

The straight inner part of the jet flares at approximately the position where it crosses
the contact discontinuity, suggesting that the jet is entraining and thermalizing
some of the hot gas as it crosses the edge.
The lobe flaring and gas entrainment were originally predicted in simulations
of \citet{loken95}.
It is not unreasonable, then, to interpret the fact that jets crossing density
edges can become partially disrupted and inflate
to form inner radio lobes
as a consequence of observed signatures of interactions
between an AGN and its surrounding medium.
The tunnel and the filaments around the lobe demonstrate that the jets
do not simply propagate losslessly through the ICM, but that even a small amount of
swept up gas isn't sufficient to completely disrupt the jets.  In fact,
entrainment of external material may play a central role in
the evolution of jets from nucleus to hot spot/termination.
Future deeper \textit{Chandra} observations of this
group would facilitate a more detailed study of the sloshing interface and
of the entrained material along the southern inner jet and lobe.


\section{Acknowledgements}

DVL thanks R.~Johnson for many fruitful conversations and
is grateful to N. P.~Lee for repeated help with the ``crude'' temperature map.
Support for this work was provided by the National Aeronautics and Space
Administration through \textit{Chandra} Award Number GO9-9111X
issued by the \textit{Chandra}
X-ray Observatory Center, which is operated by the Smithsonian Astrophysical
Observatory for and on behalf of the National Aeronautics Space Administration
under contract NAS8-03060.  JHC acknowledges support from the South-East Physics Network
(SEPNet).  This research has made use of software provided by the \textit{Chandra}
X-ray Center in the application packages {\sc CIAO} and Sherpa.
This research has made use of the NASA/IPAC Extragalactic Database (NED)
which is operated by the Jet Propulsion Laboratory, California Institute of
Technology, under contract with NASA.
This research has made use of NASA's Astrophysics Data System.



\clearpage

\end{document}